\documentclass[lettersize,journal]{IEEEtran}
\usepackage{amsmath,amsfonts}
\usepackage{algorithmic}
\usepackage{algorithm}
\usepackage{array}
\usepackage[caption=false,font=footnotesize,labelfont=footnotesize,textfont=footnotesize]{subfig}
\usepackage[
pdftex,
pdftitle={RHINO-MAG: Recursive H-Field Inference based on Observed Magnetic Flux under Dynamic Excitation},
pdfauthor={Hendrik Vater, Oliver Schweins, Lukas Hölsch, Wilhelm Kirchgässner, Till Piepenbrock, Oliver Wallscheid},
pdfcreator={Hendrik Vater, Oliver Schweins, Lukas Hölsch, Wilhelm Kirchgässner, Till Piepenbrock, Oliver Wallscheid},
linktoc=all,            
colorlinks=false,       
pdfborder={0 0 0},      
breaklinks,             
bookmarks,              
plainpages=false,       
pdfpagelabels,          
hypertexnames=false     
]{hyperref}
\usepackage{orcidlink}
\usepackage{textcomp}
\usepackage{stfloats}
\usepackage{url}
\usepackage{verbatim}
\usepackage{graphicx}
\usepackage{tabularx}
    \newcolumntype{L}{>{\raggedright\arraybackslash}X}
\usepackage{enumitem}
\usepackage{xcolor}
\usepackage{bm}
\usepackage{blindtext}
\usepackage{nicefrac}
\usepackage{booktabs}
\usepackage{siunitx}
\usepackage[acronym]{glossaries}

\usepackage{cite}

\usepackage{xcolor}
\usepackage{etoolbox}

\newtoggle{showchanges}
\togglefalse{showchanges}
\newcommand{\change}[1]{%
  \iftoggle{showchanges}{\textcolor{red}{#1}}{#1}%
}
\newcommand{\changetablecolor}{\iftoggle{showchanges}{\color{red}}{\color{black}}}

\makeatletter
\newcommand{\redkeys}[1]{%
  \@for\@tempkey:=#1\do{%
    \expandafter\gdef\csname isred@\@tempkey\endcsname{1}%
  }%
}
\redkeys{Chauhan2026,Zou2026,Wang2026,JaxEcosystem2020,Kidger2021equinox}

\let\origbibitem\bibitem
\renewcommand{\bibitem}[1]{%
  \@ifundefined{isred@#1}%
    {\color{black}}%
    {\iftoggle{showchanges}{\color{red}}{\color{black}}}%
  \origbibitem{#1}%
}
\makeatother

\usepackage{pifont}
\newcommand{\cmark}{\ding{51}}%
\newcommand{\xmark}{\ding{55}}%

\makeglossaries

\newacronym{pfc}{PFC}{power factor correction}
\newacronym{fem}{FEM}{finite element method}

\newacronym{rms}{RMS}{root mean squared}
\newacronym{sre}{SRE}{sequence relative error}
\newacronym{nere}{NERE}{normalized energy relative error}
\newacronym{lstm}{LSTM}{long short-term memory}
\newacronym{rnn}{RNN}{recurrent neural network}
\newacronym{gru}{GRU}{gated recurrent unit}
\newacronym{mc1}{MC1}{MagNet Challenge 2023}
\newacronym{mc2}{MC2}{MagNet Challenge 2025}
\newacronym{ja}{JA}{Jiles-Atherton}
\newacronym{llg}{LLG}{Landau-Lifshitz-Gilbert equation}
\newacronym{ode}{ODE}{ordinary differential equation}
\newacronym{pde}{PDE}{partial differential equation}
\newacronym{ml}{ML}{machine learning}

\newacronym{lstm-p}{LSTM-P}{LSTM with direct prediction}
\newacronym{gru-p}{GRU-P}{GRU with direct prediction}
\newacronym{gru-m}{GRU-M}{magnetization GRU}
\newacronym{gru-l}{GRU-L}{GRU parameterizes linear model}
\newacronym{p-gru-p}{P-GRU-P}{Pretrained GRU with direct prediction}
\newacronym{gru-v}{GRU-V}{vector field GRU}
\newacronym{gru-jadp}{GRU-JADP}{GRU directly parameterizes JA}
\newacronym{pinn-ja}{PINN-JA}{physics-informed neural network with JA regularization}
\newacronym{pinn}{PINN}{physics-informed neural network}

\newacronym{mse}{MSE}{mean squared error}
\newacronym{mae}{MAE}{mean absolute error}
\newacronym{wce}{WCE}{worst-case error}

\begin{document}

\title{RHINO-MAG: Recursive H-Field Inference based on Observed Magnetic Flux \change{Density} under Dynamic Excitation}

\author{Hendrik Vater \orcidlink{0009-0005-0654-8741}, Oliver Schweins \orcidlink{0009-0005-3246-0487},
Lukas Hölsch \orcidlink{0009-0006-4112-2444}, Wilhelm Kirchgässner \orcidlink{0000-0001-9490-1843}, \\ Till Piepenbrock \orcidlink{0009-0008-9483-4052}, and Oliver Wallscheid \orcidlink{0000-0001-9362-8777}, \IEEEmembership{Senior Member, IEEE}
\thanks{Hendrik Vater, Lukas Hölsch, and Oliver Wallscheid are with the Chair of Interconnected Automation Systems, University of Siegen, 57076 Siegen, Germany (e-mail: \{hendrik.vater, lukas.hoelsch, oliver.wallscheid\}@uni-siegen.de).}
\thanks{Oliver Schweins and Till Piepenbrock are with the Department of Power Electronics and Electrical Drives, Paderborn University, 33098 Paderborn, Germany (e-mail: oliverjs@mail.uni-paderborn.de, piepenbrock@lea.uni-paderborn.de).}
\thanks{Wilhelm Kirchgässner is with Beckhoff Automation, 33415 Verl, Germany (e-mail: wk.research@web.de).}

\thanks{Accepted for publication in IEEE Transactions on Power Electronics. \\© 2026 IEEE.  Personal use of this material is permitted.  Permission from IEEE must be obtained for all other uses, in any current or future media, including reprinting/republishing this material for advertising or promotional purposes, creating new collective works, for resale or redistribution to servers or lists, or reuse of any copyrighted component of this work in other works.}
}

\markboth{}%
{Shell \MakeLowercase{\textit{et al.}}: A Sample Article Using IEEEtran.cls for IEEE Journals}

\IEEEpubid{}

\glsdisablehyper
\maketitle

\begin{abstract}
Driven by the MagNet Challenge 2025 (MC2), increased research interest is directed towards modeling transient magnetic fields within ferrite materials.
An accurate time-resolved and temperature-aware H-field prediction is essential for optimizing magnetic components in applications with quasi-stationary / non-stationary excitation waveforms.
Within the scope of this investigation, a selection of model architectures with varying degrees of physically motivated structure is compared.
Based on a Pareto investigation, a rather black-box gated recurrent unit (GRU) model structure with a graceful initialization setup is found to offer the most attractive model size vs. model accuracy trade-off \change{in the small-model regime}, while the \change{examined} physics-inspired models performed worse.
For a GRU-based model \change{architecture} with only $325$ parameters \change{(trained separately per material)}, an average sequence relative error of $8.02 \, \%$ and an average normalized energy relative error of $1.07 \, \%$ across five different materials are achieved on unseen test data. With this excellent parameter efficiency, the proposed model won the first place in the performance category of the MC2.
\end{abstract}

\begin{IEEEkeywords}
power magnetics, machine learning, dynamic system modeling
\end{IEEEkeywords}

\section{Introduction}
\label{sec:introduction}

Magnetics are usually one of the bulkier parts of modern electronics and are responsible for considerable losses~\cite{Muhlethaler2012Diss}. The continuously growing demand for greater efficiency, miniaturization, and sustainability places increasing emphasis on optimized designs for magnetic components.
The optimization, in turn, requires accurate modeling and understanding of the electromagnetic behavior and corresponding core losses of the magnetic material.
The modeling is challenging because the waveforms resulting from different inputs exhibit varying degrees of hysteresis and saturation that depend on the type of material, the core temperature, and the shape, amplitude and frequency of the input waveform~\cite{Muhlethaler2012Diss, Stenglein2021Diss, Stenglein2021, Li2023}. 
This, paired with the lack of a satisfactory first-principles model for the underlying processes, has led commonly utilized models to be only accurate approximations for specific operation conditions~\cite{Li2023}. While physical first-principles models for the micromagnetic material behavior exist,
a transfer of the microscopic physical framework to macroscopic magnetic effects and to the prediction of magnetic component behavior is insufficiently investigated.

\subsection{Related work}

\IEEEpubidadjcol

For steady-state operation (e.g., in a DC-DC converter), the usual process to estimate core losses is to employ specific approximation equations, e.g., in the form of the Steinmetz equation~\cite{Steinmetz1984, Steinmetz1894}, its various improvements~\cite{Albach1996, Lancarotte2001, Venkatachalam2002, Muhlethaler2011, Muhlethaler2011improved}, and more recent developments such as the Stenglein method~\cite{Stenglein2021}.

While these steady-state core loss models are valid approximations and very effective tools under the appropriate conditions, it has been shown in the context of the \gls{mc1}, that they can reach their limits when faced with extensive and varied real-world data~\cite{Kirchgassner2024, Chen2024}.
In \gls{mc1}, this was addressed by encouraging the development of data-driven material models that predict power losses based on waveform, frequency, and temperature in ferrites in steady-state operation~\cite{Chen2024}.
Much progress towards general and accurate models for core loss prediction in steady-state has been made utilizing the tools from \gls{ml} in the context of the challenge (see, e.g.,\cite{Li2023, Chen2024, Kirchgassner2024, Huang2024, Zhang2024}).

Beyond steady-state operation, there is a vast class of applications in which quasi-stationary or semi-arbitrary excitation of the material is performed due to switched waveforms, e.g., \gls{pfc} circuits, electric motor drives, and power amplifiers~\cite{Kwon2025}.
There are simple models based on the Steinmetz equation utilizing the superposition of the major and minor loop core losses for prediction of the core losses in quasi-stationary operation of \gls{pfc} circuits~\cite{Muhlethaler2012Diss, Jacoboski2017}.
However, these concepts cannot easily be extended to semi-arbitrary excitation conditions where the frequencies for major and minor loops constantly change during operation.
For instance, in an electric drive, the rotation frequency may change during operation, making the superposition inaccurate.
Additionally, superposition inherently assumes linearity, which is a simplifying assumption for the nonlinear material behavior. Time-resolved modeling allows for more accurate consideration of the nonlinearity.
Furthermore, it enables modeling of transients and time-domain circuit simulation~\cite{Li2023}.

Classical approaches for predicting time-resolved hysteretic material behavior include the Preisach~\cite{Preisach1935} and \gls{ja}~\cite{Jiles1986} models, both of which are in essence phenomenological approximations, i.e., models that aim to mathematically describe observed data trajectories and are inspired by physical ideas but are not actually faithful to a complete physical framework~\cite{Moree2023}.

In practice, the \gls{ja} model has been used, e.g., for the estimation of core losses in electric machines with non-sinusoidal excitations~\cite{Steentjes2016, Colombo2025} and for modeling of the material behavior in \gls{fem} simulations~\cite{Gyselinck2004, Guerin2017, Benabou2003}.
Some issues have been identified with the initial formulation of the model.
There is a risk of a non-physical negative differential permeability~\cite{Ma2022} and, furthermore,
non-closure of minor loops and consequent drifting of the estimate are known problems when the excitation is not a closed loop~\cite{Moree2023}.
Multiple extensions of the \gls{ja} model have been proposed that aim to improve the prediction quality including more precise prediction of minor loops~\cite{Benabou2008, Leite2009, Kokornaczyk2014, Hamel2022}.
However, most of the existing applications of the \gls{ja} model are limited to a narrow range of operating conditions.
Some first approaches with adaptable parameters have been proposed in order to allow more general application (see, e.g.,~\cite{Hussain2017, Ma2022}).

The Preisach model has been used for iron loss estimations with different excitation waveforms (e.g.,\cite{Darabi2007, Zhao2020, Li2024}) and also has been compared to the \gls{ja} model in terms of predictive performance for DC-biased hysteresis behavior~\cite{Li2024}. 
The Preisach model was found to be more accurate, especially for increasing DC-bias, but also required more computational effort~\cite{Li2024}.
In~\cite{Liu2023}, it was found that the proposed inverse Preisach was more accurate and required less computational effort compared to the inverse \gls{ja} model. However, a closed form for the underlying Everett function was utilized in~\cite{Liu2023}, reducing the flexibility of the Preisach model.

Ferrite materials are made up of an arbitrarily oriented and shaped grain structure on the micrometer scale that heavily dominates the macroscopic magnetic behavior.
Physical models based on micromagnetic particle interactions such as the \gls{llg}~\cite{Landau1935, Gilbert2004} operate at the nanometer scale.
Transferring those to real-world magnetic material probes with heterogeneous, defective structures remains an open question.

In summary, a first-principles model to explain magnetic hysteresis that is reasonably applicable on a macroscopic scale is missing~\cite{Li2023}.
Existing phenomenological, data-driven models \change{have only been applied to distinct, narrowly defined subproblems and have not, to the best knowledge of the authors, been compared under a common benchmark task spanning various operating conditions.} This makes it difficult to determine which model types are most \change{promising} for varied and semi-arbitrary excitations.

\subsection{MagNet Challenge 2025}
\label{subsec:mc2_data_set}

Driven by the gap in physical understanding, modeling capabilities, and heterogeneous data for time-resolved magnetic material behavior and further motivated by the success of data-driven models for steady-state operation in \gls{mc1}, \gls{mc2}~\cite{Kwon2025} emphasizes transient magnetic material behavior.
The goal is to reconstruct and understand the time-resolved trajectories of the magnetic field~$H$, given histories of magnetic flux density~$B$ and other relevant operation information to enable a better understanding of the physical behavior and more accurate predictions of core losses.
For this purpose, a corresponding data set\footnote{The publication corresponding to the data set is~\cite{Kwon2025} and the accompanying repository can be found at \url{https://github.com/PaulShuk/MagNetX}.} is provided by the competition hosts~\cite{Kwon2025}.
Additionally, first data-driven models have been proposed by the competition hosts for such time-resolved, transient modeling of magnetic material behavior in~\cite{Wang2024, Wang2025} using \gls{lstm}-based \glspl{rnn}.

\begin{table}[t]
    \caption{\change{MagNet Challenge 2025 data set overview}}
    \centering
    \addtolength{\tabcolsep}{-0.1em}
    \begin{tabularx}{\linewidth}{>{\changetablecolor}l >{\changetablecolor}c >{\changetablecolor}l}\toprule
    Name & Symbol & Value\\
    \midrule
    Pretest materials & $m_1 \in M_1$ & \begin{minipage}[t]{\linewidth}
        $M_1 = \{ 
            \mathrm{3C90}, \mathrm{3C94}, \mathrm{3E6}, \mathrm{3F4}, \mathrm{77}, \mathrm{78},  \\ \mathrm{N27},\mathrm{N30}, \mathrm{N49}, \mathrm{N87}\}$
    \end{minipage} \\
    Pretest \#$H$ meas. & & $\sim870 \times 10^6$ data points \\
    Eval.~materials & $m_2 \in M_2$ & \begin{minipage}[t]{\linewidth}
        $M_2 = \{\mathrm{3C92}, \mathrm{3C95}, \mathrm{FEC007}, \\ \mathrm{FEC014}, \mathrm{T37}\}$
    \end{minipage} \\
    Eval.~\#$H$ meas. & & \begin{minipage}[t]{\linewidth}
        $\sim226 \times 10^6$ training data points \\
        $\sim 5.6 \times 10^6 $ testing data points
    \end{minipage} \\
    Switching freq. & $f_\mathrm{sw}$ & $\{50, 80, 125, 200, 320, 500, 800\} \, \SI{}{\kilo\hertz}$ \\
    Temperature & $\vartheta$ & $\{25, 50, 70\} \, \SI{}{\celsius}$ \\
    Sampling period & $\tau$ & $\nicefrac{1}{\SI{16}{\mega\hertz}} = \SI{62.5}{\nano\second}$ \\
    \bottomrule
    \end{tabularx}
    \label{tab:data_set_parameters}
\end{table}

\change{An overview of the \gls{mc2} data set is provided in Tab.~\ref{tab:data_set_parameters}.
Measurement data is provided for $15$ different ferrite materials. For $10$ of these materials, all of the available data for each material $\mathcal{D}^{(m_1)}$ was fully disclosed in a pretest phase so that the competitors could investigate different models.
In an evaluation phase,} the training data $\mathcal{D}^{(m_2)}_\mathrm{train}$ for $5$ further materials $m_2 \in M_2$ was provided, while the corresponding testing data $\mathcal{D}^{(m_2)}_\mathrm{test}$ was withheld. The names of the materials in $M_2$ were anonymized during the challenge. 
The competitors were tasked to train their models on $\mathcal{D}^{(m_2)}_\mathrm{train}$ and the model performance was evaluated on $\mathcal{D}^{(m_2)}_\mathrm{test}$ by the competition hosts.

Each material data set contains a collection of measurement sequences.
Each sequence was generated by exciting the material with a randomly varying duty cycle resulting in quasi-transient waveforms for $B$ and $H$ (see~\cite{Kwon2025} for details on the data acquisition and properties). 

\begin{figure}[htbp]
    \centering
    \includegraphics[width=\columnwidth]{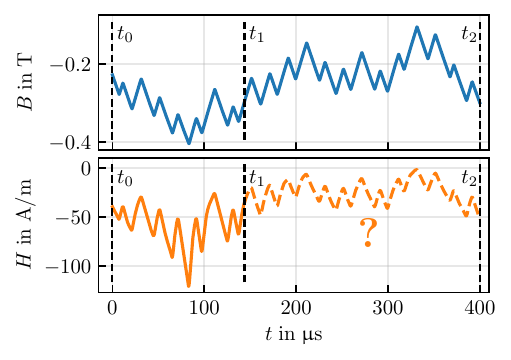}%
    \caption{Exemplary prediction task for $H$ based on available $B$.}
    \label{fig:prediction_task_example}
\end{figure}

An exemplary trajectory prediction task is provided in Fig.~\ref{fig:prediction_task_example}.
The sampling index $k \in \mathbb{N}$ is calculated as $k = \nicefrac{t}{\tau}$ based on the measurement time $t$ and the sampling period $\tau$.
A single measurement sequence starts with $k=0$ and ends at $k_3 \in \mathbb{N}_+$. Note that $k_3$ varies depending on $f_\mathrm{sw}$~\cite{Kwon2025}.
A prediction sequence starts with $k_0 \in \mathbb{N}$, while $k_1 \in \mathbb{N}_+$ is the first time index at which no more $H$ information is available and $H$ needs to be predicted until $k_2 \in \mathbb{N}_+$. 
The sampling indices must satisfy $0 \leq k_0 < k_1 < k_2 \leq k_3$.

The prediction accuracy of a model is measured by the \gls{sre} score
\begin{align}
    \mathrm{SRE} = \sqrt{\frac{\sum_{k=k_1}^{k_2} \left( \hat{H}_k - H_k \right)^2
    }{
    \sum_{k=k_1}^{k_2} H_k^2 }},
    \label{eq:SRE}
\end{align}
and the accuracy in terms of core loss estimation by the \gls{nere} score
\begin{align}
    \mathrm{NERE} = \frac{\sum_{k=k_1}^{k_2} \left( \Delta B_k \hat{H}_k \right) - \sum_{k=k_1}^{k_2} \left( \Delta B_k H_k \right)
    }{
    \sum_{k=0}^{k_3} \left( \Delta B_k H_k \right) },
    \label{eq:NERE}
\end{align}
on unseen test data, where $\hat{H}$ is the model's prediction for $H$.
\change{Other participants of the \gls{mc2} have reported their model architectures (e.g., from team 'Hefei'~\cite{Wang2026}, 'Tianjin'~\cite{Zou2026}, and 'Georgia Tech'~\cite{Chauhan2026}). A comparison of results reported by the \gls{mc2} organizing committee is provided as part of the extended Pareto front investigation in Section~\ref{subsec:extended_pf}.}

\subsection{Contribution}
\label{subsec:contribution}

\change{Against the above background, this work makes the following contributions:}
\begin{enumerate}
    \item Design of a parameter-efficient \gls{gru} model archetype for time-resolved modeling of magnetic material behavior with an intuitive and effective concept for warmup of the model's hidden state. 
    \item Extensive comparison of a selection of model archetypes for a variety of model sizes with respect to their prediction accuracy in order to provide a Pareto front and, therefore, a basis of comparison for further magnetic transient modeling approaches.
    \item \change{Ablation studies for the utilized training loss function and for the warmup procedure to examine their individual contribution to the model performance.}
    \item Open-source release of the RHINO-MAG modeling framework containing all of the presented models.\footnote{All code used for the presented results is openly available at \url{https://github.com/upb-lea/RHINO-MAG}.} The framework is implemented using the JAX\footnote{Documentation for the JAX library: \url{https://docs.jax.dev/en/latest/}.} Python library for seamless utilization of accelerator hardware (GPUs, TPUs) and just-in-time (JIT) compilation.
\end{enumerate}

With only $325$ parameters and an average \gls{sre} score of $8.02 \, \%$ and an average \gls{nere} score of $1.07 \, \%$ on unseen \change{quasi-transient} test data, the proposed \gls{gru-p} model \change{architecture} won the first place in the performance track in the \gls{mc2}, showing an excellent accuracy-per-parameter efficiency \change{especially in the small-model regime}.
\change{Note that the models are trained on a per material basis and inference is performed on data that was held out from the training data set.}

\section{General modeling setup}

The modeling task is to estimate the scalar component of the magnetic field along the excitation and measurement direction $\hat{H}_k$ with $k \in \left[k_1, k_2\right]$ based on the previously observed magnetic field $H_k$ with $k \in \left[k_0, k_1\right)$, the magnetic flux density $B_k$ with $k \in \left[k_0, k_2\right]$, and the core temperature $\vartheta$.
That is, a model of the form
\begin{align}
    \hat{\bm{H}}_{k_1:k_2} = \bm{\mathcal{M}}(\bm{B}_{k_0:k_2}, \bm{H}_{k_0:k_1-1}, \vartheta)
    \label{eq:model_base_equation}
\end{align}
is sought after (see~\cite{Wang2025}). 
The notation $\bm{H}_{k_0:k_1-1}$ describes a time series of field values
\begin{equation}
    \bm{H}_{k_0:k_1-1} = \begin{bmatrix}
        H_{k_0},  H_{k_0+1}, \dots  H_{k_1-1}
    \end{bmatrix},
\end{equation}
and it is used analogously for all other quantities.

In the following, concepts that are shared among most of the considered models are presented.
Whenever the training setup for a model varies from the described form, it is noted explicitly in the respective section of the model in Sec.~\ref{sec:model_types}.

\subsection{Feature Engineering}
\label{subsec:FE}

The term feature engineering encompasses all preprocessing, normalization, and derivation of additional features in an observational data set.
At this point it should be noted that it is possible to extract the switching frequency $f_\mathrm{sw}$ from the considered data set. 
This has explicitly not been done \change{in order to keep the model architecture applicable to excitation signal shapes beyond the quasi-transient, fixed-frequency waveforms present in the training data. The generalizability of this is not directly verified in the scope of this work.}

\subsubsection{Normalization}
The normalization is performed by first computing the maximum absolute value for the raw magnetic field, magnetic flux density, and temperature values $H_\mathrm{max}$, $B_\mathrm{max}$, and $\vartheta_\mathrm{max}$, respectively, for each material \change{training} data set independently. The normalization is then performed by simple division: 
\begin{align}
    \tilde{z} = \frac{z}{z_\mathrm{max}} \quad \text{with} \quad z\in \{H, B, \vartheta\}.
\end{align}

\subsubsection{Features}

The featurized input matrix\footnote{As $k_1-1 < k_2$, the vector of known $H$ values $\bm{H}_{k_0:k_1-1}$ is shorter than the width of the matrix $\bm{X}_{k_0:k_2}$.
Therefore, $\bm{H}_{k_0:k_1-1}$ is processed independently of $\bm{X}_{k_0:k_2}$ and not considered part of the feature matrix.} is built as
\begin{align}
    \bm{X}_{k_0:k_2} = \begin{bmatrix}
        \tilde{B}_{k_0} & \tilde{B}_{k_0 + 1} & \dots & \tilde{B}_{k_2} \\
        \Delta \tilde{B}_{k_0} & \Delta \tilde{B}_{k_0 + 1} & \dots & \Delta \tilde{B}_{k_2} \\
        \Delta^2 \tilde{B}_{k_0} & \Delta^2 \tilde{B}_{k_0 + 1} & \dots & \Delta^2 \tilde{B}_{k_2} \\
        \tilde{\vartheta} & \tilde{\vartheta} & \dots & \tilde{\vartheta} \\
    \end{bmatrix},
\end{align}
where $\Delta \tilde{B}$ and $\Delta^2 \tilde{B}$ are finite-difference-based approximations of the first and second derivatives, respectively. Those are associated with the switching behavior of the feeding inverter during measurements and are considered physics-inspired input signals helping the model learn the switching-induced transients of the $BH$-curve. 
While the sampling time of $\tau = \SI{62.5}{\nano\second}$ is known, $\tau^\prime = \SI{1}{\second}$ is set in the finite-difference calculation for improved numerical stability. This can be interpreted as a normalization of the time.
Further features, such as dynamic averaging of signals, were tested but did not improve the model performance.

\subsection{Training cost functions}

An adapted \gls{rms} error was utilized as the loss function:
\begin{align}
    \mathcal{L}_{\mathrm{\change{a}RMSE}} &= \sqrt{\frac{ \sum_{k=k_1}^{k_2} \left[\left( \tilde{H}_k - \hat{\tilde{H}}_k\right)^2 \cdot \left| \tilde{B}_k - \tilde{B}_{k-1} \right|  \right]}{k_2 - k_1 + 1}}.
    \label{eq:rmse_loss}
\end{align}
Additionally, the loss is normalized with the \gls{rms} value of the full sequence $\bm{H}_{0:k_3}$ from which $\tilde{\bm{H}}_{k_0:k_2}$ and $\tilde{\bm{B}}_{k_0:k_2}$ are sampled. 
Specifically, the weighted loss
\begin{align}
    \mathcal{L}^\prime_{\mathrm{\change{a}RMSE}} = \mathcal{L}_{\change{a}\mathrm{RMSE}} \cdot H_{\mathrm{max}} \cdot  \left( \frac{1}{k_3 + 1} \sum_{k=0}^{k_3} H_k^2 \right)^{-\frac{1}{2}}
    \label{eq:weighted_rmse_loss}
\end{align}
is backpropagated to obtain the gradient with respect to model parameters $\bm{\theta}$.
The cost function~\eqref{eq:rmse_loss} combines elements of both the \gls{sre} and \gls{nere} metrics. The quadratic tracking error on $H$, adopted from \gls{sre}, promotes accurate $H$ prediction, while the pointwise errors are weighted by the change in $B$ to incorporate the energy-related considerations inherent in \gls{nere}.
The normalization by the \gls{rms} value of the full sequence in~\eqref{eq:weighted_rmse_loss} prevents sequences with lower excitation frequencies from being overweighted, as their heavily saturated regions give rise to disproportionately large $H$ values.

\subsection{Model structure overview}

\begin{figure}[htbp]
    \centering
    \includegraphics[width=\columnwidth]{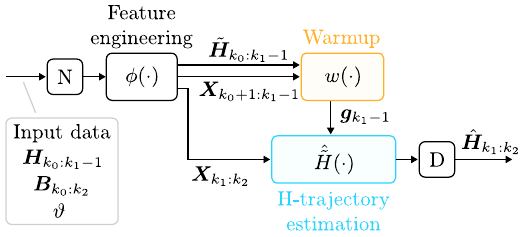}%
    \caption{Overview of the model structure. 'N' is a normalization operation, while 'D' is the denormalization.}
    \label{fig:overview}
\end{figure}

An overview of the proposed model structure~\eqref{eq:model_base_equation} is given in Fig.~\ref{fig:overview}.
The raw input data is normalized, featurized, and then split into two parts. The first part consists of $\bm{X}_{k_0+1:k_1-1}$ and $\tilde{\bm{H}}_{k_0:k_1-1}$ and is used for the warmup of the hidden state $\bm{g}$ of the model.
The second part is $\bm{X}_{k_1:k_2}$ and it is used together with the warmed-up hidden state $\bm{g}_{k_1-1}$ for the estimation of the normalized future field values $\hat{\tilde{\bm{H}}}_{k_1:k_2}$, which are then denormalized for the final predicted sequence $\hat{\bm{H}}_{k_1:k_2}$.
An in-depth view of the warmup and prediction process is provided for the \gls{gru-p} model type in Fig.~\ref{fig:detailed_process}.

\subsection{Parameter updates}

Each material-specific model is trained for multiple epochs over the training dataset.
For a single epoch, the training data is split into subsequences of the same length $l \in \mathbb{N}_+$, and then $b \in \mathbb{N}_+$ subsequences are randomly grouped to form a mini-batch
\begin{equation}
    \mathcal{B} = \left\{\mathbf{B}_{k_0:k_2} \in \mathbb{R}^{b \times l}, \mathbf{H}_{k_0:k_2} \in \mathbb{R}^{b \times l}, \bm{\vartheta} \in \mathbb{R}^{b \times 1} \right\}.
\end{equation}
This way, a list of mini-batches is precomputed for the whole training data set.
These mini-batches are iterated over once and for each mini-batch the model $\bm{\mathcal{M}}$ is used to perform a prediction  $\hat{\mathbf{H}}_{k_1:k_2}$. 
The training loss~\eqref{eq:weighted_rmse_loss} is evaluated for $\hat{\mathbf{H}}_{k_1:k_2}$ and $\mathbf{H}_{k_1:k_2}$, the gradient with respect to the model parameters $\nabla_{\bm{\theta}} \mathcal{L}^\prime_{\mathrm{\change{a}RMSE}}$ is computed via backpropagation, and the parameter update is computed based on it.
Finally, a new set of sequences and ordering is computed for the following epoch.

\section{Model types}
\label{sec:model_types}

A collection of model types with varying degrees of physical motivation has been investigated in the context of this work.
Here, the structures of these models are presented.
An overview of all provided models is given in Sec.~\ref{subsec:model_overview}.

\subsection{GRU Variants}

\begin{figure*}[htbp]
    \centering
    \includegraphics[width=\textwidth]{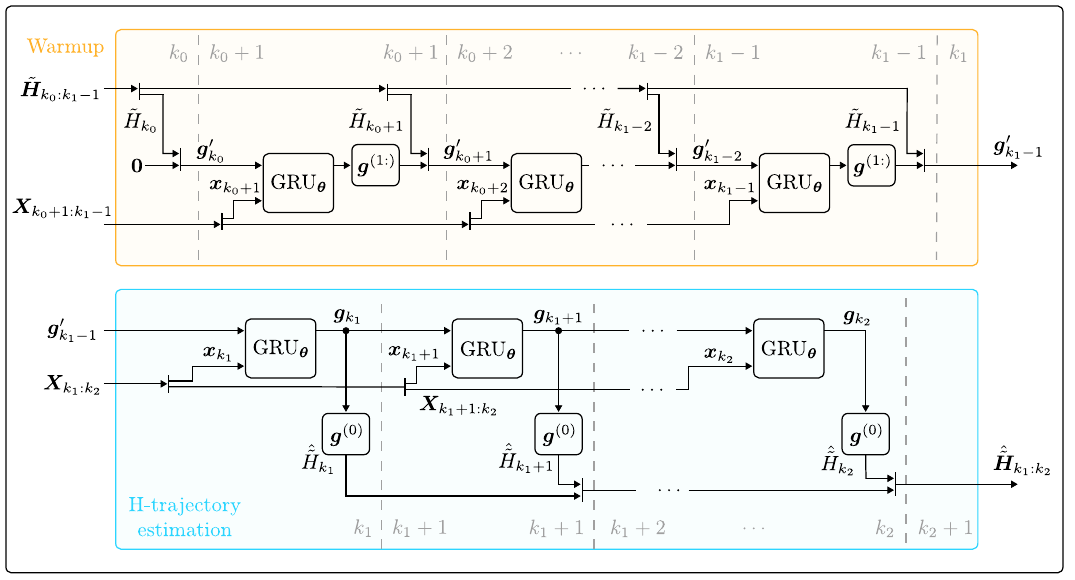}%
    \caption{Detailed view of model warmup and $H$-trajectory estimation for the \gls{gru-p}. The superscript numbers resemble specific indexing of a vector (in Python style).}
    \label{fig:detailed_process}
\end{figure*}

To reliably predict the magnetic field $H$ based on the magnetic flux $B$, the hysteresis of the material should be considered in the model.
The model needs to account for the hysteretic behavior of the material which, in essence, is a hidden state of the problem.
Therefore, it is intuitive to choose a model from the class of \glspl{rnn}. First, a \gls{gru} cell is considered~\cite{Kidger2021}
\begin{equation}
\begin{aligned}
    \bm{z}_k &= \sigma\left( \bm{W}_\mathrm{z} \bm{x}_k + \bm{b}_\mathrm{z} + \bm{U}_\mathrm{z} \bm{g}_{k-1} \right)\\
    \bm{r}_k &= \sigma\left( \bm{W}_\mathrm{r} \bm{x}_k + \bm{b}_\mathrm{r} + \bm{U}_\mathrm{r} \bm{g}_{k-1} \right)\\
    \tilde{\bm{g}}_{k} &= \tanh \left( \bm{W} \bm{x}_k + \bm{b} + \bm{r}_k \odot  \left(\bm{U} \bm{g}_{k-1} + \bm{b}_\mathrm{n} \right)\right) \\
    \bm{g}_k &= \tilde{\bm{g}}_{k} + \bm{z}_k \odot \left( \bm{g}_{k-1} - \tilde{\bm{g}}_{k} \right), 
\end{aligned}
\label{eq:GRU_cell_eq}
\end{equation}
where $\bm{g}_k \in \mathbb{R}^{d_\mathrm{g}}$ is the hidden state, the input is $\bm{x}_k \in \mathbb{R}^{d_\mathrm{x}}$, the 
update gate parameters are $\bm{W}_\mathrm{z} \in \mathbb{R}^{d_\mathrm{g} \times d_\mathrm{x}}$, $\bm{b}_\mathrm{z} \in \mathbb{R}^{d_\mathrm{g}}$, $\bm{U}_\mathrm{z} \in \mathbb{R}^{d_\mathrm{g} \times d_\mathrm{g}}$, the reset gate parameters are $\bm{W}_\mathrm{r} \in \mathbb{R}^{d_\mathrm{g} \times d_\mathrm{x}}$, $\bm{b}_\mathrm{r} \in \mathbb{R}^{d_\mathrm{g}}$, $\bm{U}_\mathrm{r} \in \mathbb{R}^{d_\mathrm{g} \times d_\mathrm{g}}$, the remaining parameters are $\bm{W} \in \mathbb{R}^{d_\mathrm{g} \times d_\mathrm{x}}$, $\bm{b} \in \mathbb{R}^{d_\mathrm{g}}$, $\bm{U} \in \mathbb{R}^{d_\mathrm{g} \times d_\mathrm{g}}$, $\bm{b}_\mathrm{n} \in \mathbb{R}^{d_\mathrm{g}}$, and $\sigma$ is the sigmoid activation function.
In the following, \eqref{eq:GRU_cell_eq} is written as
\begin{equation}
    \bm{g}_k = \mathrm{GRU}_{\bm{\theta}} \left(\bm{x}_k, \bm{g}_{k-1} \right),
    \label{eq:GRU_shorthand}
\end{equation}
where $\bm{\theta}$ are the tunable parameters of the cell.

\subsubsection{GRU with direct prediction (GRU-P)}
\label{subsubsec:GRU_direct_pred}

The warmup process (upper part of Fig.~\ref{fig:detailed_process}) aims to utilize the portion of the data where both $H$ and $B$ are known.
For this model variant, the first hidden state is created by concatenating the first normalized field value with zeros
\begin{equation}
    \bm{g}^\prime_{k_0} = \begin{bmatrix}\tilde{H}_{k_0} \\ \bm{0}\end{bmatrix}.
\end{equation}
The featurized input sequence 
\begin{equation}
    \bm{X}_{k_0 +1:k_1-1} = \left[\bm{x}_{k_0 + 1}, \bm{x}_{k_0 + 2}, \dots, \bm{x}_{k_1-1} \right]
\end{equation}
is then fed sequentially into~\eqref{eq:GRU_shorthand}. The corresponding hidden state is created analogously to the first hidden state, where the first element is always set to the true normalized magnetic field value, while the remaining elements of the vector are filled with the hidden state produced by \eqref{eq:GRU_shorthand} in the last iteration\footnote{The notation $\bm{z}^{(s:)}$ with $\bm{z} \in \mathbb{R}^{d_{\bm{z}}}$ and $s \in \{0, 1, ..., d_{\bm{z}}\}$ is used as a shorthand for \mbox{$\bm{z}^{(s:)} = \begin{bmatrix}z_s & z_{s+1} & \dots & z_{d_{\bm{z}}-1} \end{bmatrix}^\top \in \mathbb{R}^{d_{\bm{z}} - s}$}. Additionally, \mbox{$\bm{z}^{(s)} = z_s\in\mathbb{R}$} and \mbox{$\bm{z}^{(:s)} = \begin{bmatrix} z_0 & z_1 \dots & z_{s-1} \end{bmatrix}^\top \in \mathbb{R}^s$} are defined.}
\begin{equation}
    \bm{g}^\prime_{k} = \begin{bmatrix}\tilde{H}_{k} \\ \bm{g}^{(1:)}_{k}\end{bmatrix}.
    \label{eq:H_past_injection}
\end{equation}

The final hidden state of the warmup process $\bm{g}^\prime_{k_1-1}$ is then given to the $H$-trajectory estimation process (lower part of Fig.~\ref{fig:detailed_process}).
There, the featurized input sequence 
\begin{equation}
    \bm{X}_{k_1:k_2} = \left[\bm{x}_{k_1}, \bm{x}_{k_1 + 1}, \dots, \bm{x}_{k_2} \right]
\end{equation}
is fed sequentially into~\eqref{eq:GRU_shorthand}, while the hidden state from the last iteration is always looped back.
In each iteration, the first element (index $0$) of the hidden state vector is used as the normalized prediction for the magnetic field, i.e.,
\begin{equation}
    \hat{\tilde{H}}_k = \hat{\tilde{H}}_{\bm{\theta}}\left(\bm{x}_k, \bm{g}_{k-1} \right) = \bm{g}_{k}^{(0)},
    \label{eq:direct_prediction_function}
\end{equation}
which finally produces the estimated H-field trajectory $\hat{\bm{H}}_{k_1:k_2}$ after concatenation and denormalization.
\change{During training, the backpropagation continues through all components of the state $\bm{g}^{(1:)}_{k}$ that are retained and depend on the model parameters $\bm{\theta}$. As the injected $\tilde{H}_{k}$ is fully independent of the parameters, the gradient does not propagate through it.}

The main advantage of this process is that it gives the model a natural way of adapting all elements of the hidden state except the one interpreted as the $H$-field estimation to the presented data in the warmup phase without accumulation of prediction errors. 
Additionally, arbitrary warmup phase lengths (as part of the model initialization) are supported through this approach.

\subsubsection{Magnetization GRU (GRU-M)}
\label{subsubsec:magnetization_gru}
This variant works similarly to the \gls{gru-p}. 
The main difference is that it explicitly predicts the magnetization instead of directly predicting $H$
\begin{equation}
    \hat{H}_k = \hat{H}_{\bm{\theta}}\left(\bm{x}_k, \bm{g}_{k-1} \right) = \frac{{B}_k}{\mu_0} - \bm{g}_{k}^{(0)} H_{\mathrm{max}},
    \label{eq:mag_gru_pred_phys}
\end{equation}
where $\mu_0$ is the magnetic permeability in vacuum.
The idea is that the prediction function provides a comparatively direct and physically founded way for the flux density to impact the prediction. However, the computation is numerically suboptimal, as $\frac{{B}_k}{\mu_0}$ is in the range of roughly $\SI{e6}{\ampere\per\second}$ and the resulting field prediction is usually below a scale of $\SI{e2}{\ampere\per\second}$. Furthermore, it might be an issue that the permeability may introduce not only a scaling but also a time shift between $B$ and $H$, which cannot be represented with \eqref{eq:mag_gru_pred_phys}. Overall, the performance of this model is very poor and training is unstable.

For better numerical stability, another version with
\begin{equation}
    \hat{\tilde{H}}_k = \hat{\tilde{H}}_{\bm{\theta}}\left(\bm{x}_k, \bm{g}_{k-1} \right) =  \mathrm{tanh}\left(\tilde{B}_k - \bm{g}_{k}^{(0)} \right),
    \label{eq:mag_gru_pred}
\end{equation}
is tested.
The warmup for this model is done by inverting the prediction function \eqref{eq:mag_gru_pred}
\begin{equation}
    \bm{g}^\prime_{k} = \begin{bmatrix} \tilde{B}_k - \mathrm{tanh}^{-1}(\tilde{H}_{k})  \\ \bm{g}^{(1:)}_{k}\end{bmatrix}.
    \label{eq:mag_gru_warmup}
\end{equation}
The latter setup is provided as \gls{gru-m} in Sec.~\ref{sec:results}.

\subsubsection{GRU parameterizes linear model (GRU-L)}
\label{subsubsec:GRU_linear_model}
This variant utilizes the first element of the hidden state of the \gls{gru} to predict the magnetic permeability of the material. Specifically, the normalized inverse of the permeability is predicted in each step:
\begin{equation}
    \hat{\tilde{H}}_k = \hat{\tilde{H}}_{\bm{\theta}}\left(\bm{x}_k, \bm{g}_{k-1} \right) = \bm{g}_{k}^{(0)} \cdot \tilde{B}_k = \hat{\tilde{\mu}}^{-1}_k \cdot \tilde{B}_k,
    \label{eq:gtu_l_pred}
\end{equation}
where the estimate for the physical permeability can be calculated as
\begin{equation}
    \hat{\mu}_k = \frac{B_{\mathrm{max}}}{\bm{g}_{k}^{(0)}H_{\mathrm{max}}} = \hat{\tilde{\mu}}_k \frac{B_{\mathrm{max}}}{H_{\mathrm{max}}}.
\end{equation}
Analogous to the previous variants, the warmup can then be performed as the inverse of the prediction function~\eqref{eq:gtu_l_pred}
\begin{equation}
    \bm{g}^\prime_{k} = \begin{bmatrix} \tilde{H}_k\tilde{B}_k^{-1}  \\ \bm{g}^{(1:)}_{k}\end{bmatrix}.
    \label{eq:gtu_l_warmup}
\end{equation}

Similarly to the \gls{gru-m}, this setup does not incorporate the time shift characteristic of the material.
An option to circumvent this is to provide multiple past/future elements of the $B$ trajectory instead of just $B_k$, however, this places
a stronger demand on the hidden state, as each input element requires an element of the hidden state as a factor.
In the end, the \gls{gru-l} as described in \eqref{eq:gtu_l_pred} and \eqref{eq:gtu_l_warmup} is evaluated in Sec.~\ref{sec:results}.

\subsection{LSTM with direct prediction (LSTM-P)}
\label{subsec:lstm_with_direct_prediction}

An \gls{lstm} cell is defined as~\cite{Gers2000}
\begin{equation}
\begin{aligned}
    \bm{i}_k &= \sigma\left( \bm{W}_\mathrm{i} \bm{x}_k + \bm{b}_\mathrm{i} + \bm{U}_\mathrm{i} \bm{g}_{k-1} \right)\\
    \bm{f}_k &= \sigma\left( \bm{W}_\mathrm{f} \bm{x}_k + \bm{b}_\mathrm{f} + \bm{U}_\mathrm{f} \bm{g}_{k-1} \right)\\
    \bm{m}_k &= \mathrm{tanh}\left( \bm{W}_\mathrm{m} \bm{x}_k + \bm{b}_\mathrm{m} + \bm{U}_\mathrm{m} \bm{g}_{k-1} \right)\\
    \bm{o}_k &= \sigma\left( \bm{W}_\mathrm{o} \bm{x}_k + \bm{b}_\mathrm{o} + \bm{U}_\mathrm{o} \bm{g}_{k-1} \right)\\
    \bm{c}_k &= \bm{f}_k \odot \bm{c}_{k-1} + \bm{i}_k \odot \bm{m}_k \\
    \bm{g}_k &= \bm{o}_k \odot \mathrm{tanh}(\bm{c}_k) \\ 
\end{aligned}
\label{eq:LSTM_cell_eq}
\end{equation}
where $\bm{c}_k \in \mathbb{R}^{d_\mathrm{g}}$ is the cell state, $\bm{g}_k \in \mathbb{R}^{d_\mathrm{g}}$ is the hidden state, $\bm{x}_k \in \mathbb{R}^{d_\mathrm{x}}$ is the input, $\bm{W}_\mathrm{i} \in \mathbb{R}^{d_\mathrm{g} \times d_\mathrm{x}}$, $\bm{b}_\mathrm{i} \in \mathbb{R}^{d_\mathrm{g}}$, $\bm{U}_\mathrm{i} \in \mathbb{R}^{d_\mathrm{g} \times d_\mathrm{g}}$ are the input gate parameters, $\bm{W}_\mathrm{f} \in \mathbb{R}^{d_\mathrm{g} \times d_\mathrm{x}}$, $\bm{b}_\mathrm{f} \in \mathbb{R}^{d_\mathrm{g}}$, $\bm{U}_\mathrm{f} \in \mathbb{R}^{d_\mathrm{g} \times d_\mathrm{g}}$ are the forget gate parameters,  $\bm{W}_\mathrm{m} \in \mathbb{R}^{d_\mathrm{g} \times d_\mathrm{x}}$, $\bm{b}_\mathrm{m} \in \mathbb{R}^{d_\mathrm{g}}$, $\bm{U}_\mathrm{m} \in \mathbb{R}^{d_\mathrm{g} \times d_\mathrm{g}}$ are the input squashing parameters, and $\bm{W}_\mathrm{o} \in \mathbb{R}^{d_\mathrm{g} \times d_\mathrm{x}}$, $\bm{b}_\mathrm{o} \in \mathbb{R}^{d_\mathrm{g}}$, $\bm{U}_\mathrm{o} \in \mathbb{R}^{d_\mathrm{g} \times d_\mathrm{g}}$ are the output gate parameters.
In the following~\eqref{eq:LSTM_cell_eq} will be written as:
\begin{equation}
    \begin{bmatrix}
        \bm{g}_k \\ \bm{c}_k
    \end{bmatrix} = \mathrm{LSTM}_{\bm{\theta}}(\bm{x}_k, \bm{g}_{k-1}, \bm{c}_{k-1}),
    \label{eq:LSTM_shorthand}
\end{equation}
where $\bm{\theta}$ holds the tunable parameters of the \gls{lstm} cell.

The prediction and the warmup for the \gls{lstm-p} are implemented through the use of the same mechanism as \gls{gru-p} (see Sec.~\ref{subsubsec:GRU_direct_pred}):
\begin{equation}
    \hat{\tilde{H}}_k = \hat{\tilde{H}}_{\bm{\theta}}\left(\bm{x}_k, \bm{g}_{k-1}, \bm{c}_{k-1}\right) = \bm{g}_{k}^{(0)}.
    \label{eq:lstm_out}
\end{equation}
For the LSTM, it is unclear how the propagation of $\bm{c}_k$ is influenced by the warmup via~\eqref{eq:lstm_out}, compared to the \gls{gru} where fewer operations are applied to the model's single hidden state (cf.~\eqref{eq:GRU_cell_eq}).

\subsection{JA variants}
\label{subsec:ja}

While the \gls{ja} equations are phenomenological rather than first-principles equations and can be inaccurate especially regarding the temporal behavior, it is still a common tool for modeling of quasi-static hysteresis~\cite{Zirka2012}. 
In the following, multiple models that aim to incorporate the \gls{ja} formulas are presented.

\subsubsection{Basic (inverse) JA}
\label{subsubsec:basic_inverse_ja}
According to the \gls{ja} model, the change in magnetization w.r.t. changes in the magnetic field can be approximated as~\cite{Zirka2012, Zaman2016}
\begin{align}
    \frac{\mathrm{d}M}{\mathrm{d}H} = \frac{\delta_\mathrm{M} (M_\mathrm{an} - M) + ck^\prime\delta \frac{\mathrm{d}M_\mathrm{an}}{\mathrm{d}H_\mathrm{e}}}{k^\prime\delta - \alpha_\mathrm{W} \left[ \delta_\mathrm{M} (M_\mathrm{an} - M) + ck^\prime\delta \frac{\mathrm{d}M_\mathrm{an}}{\mathrm{d}H_\mathrm{e}} \right]},
    \label{eq:JA_dmdh}
\end{align}
where $\alpha_\mathrm{W}$ is related to the interaction of Weiss' domains, $a$ is a form factor, $c$ is a coefficient related to the reversibility of the domain walls, $k^\prime$ is a coefficient related to hysteresis losses, and $M_\mathrm{s}$ is the saturation magnetization~\cite{Zaman2016}.
Furthermore, $H_\mathrm{e} = H + \alpha_\mathrm{W} M$ is the effective magnetic field, $M_\mathrm{an}$ is the anhysteretic magnetization calculated as
\begin{equation}
    M_\mathrm{an} = M_\mathrm{s} \left( \mathrm{coth}\left( \frac{H_\mathrm{e}}{a} \right) - \frac{a}{H_\mathrm{e}} \right),
\end{equation}
$\delta$ is a directional parameter with~\cite{Zirka2012}
\begin{equation}
   \delta = \begin{cases}
      1 & \text{if $\frac{\mathrm{d}B}{\mathrm{d}t} > 0$}\\
      0  & \text{if $\frac{\mathrm{d}B}{\mathrm{d}t} = 0$}\\
      -1 & \text{if $\frac{\mathrm{d}B}{\mathrm{d}t} < 0$,}\\
    \end{cases}
    \label{eq:delta}
\end{equation}
and $\delta_\mathrm{M}$ is~\cite{Zirka2012}
\begin{equation}
   \delta_\mathrm{M} = \begin{cases}
      0 & \text{if $\frac{\mathrm{d}B}{\mathrm{d}t} < 0$ and $M_\mathrm{an} > M$}\\
      0 & \text{if $\frac{\mathrm{d}B}{\mathrm{d}t} > 0$ and $M_\mathrm{an} < M$}\\
      1 & \text{otherwise.}
    \end{cases}
    \label{eq:delta_M}
\end{equation}

Based on the derivations in~\cite{Zirka2012}, it follows that
\begin{equation}
    \frac{\mathrm{d}H}{\mathrm{d}t} = \frac{\mathrm{d}B}{\mathrm{d}t} \frac{1}{\mu_0} \left[ 1 - \frac{\frac{\mathrm{d}M}{\mathrm{d}H}}{1+\frac{\mathrm{d}M}{\mathrm{d}H}} \right],
    \label{eq:JA_dhdt}
\end{equation}
which is integrated using an \gls{ode} solver to provide estimates for the magnetic field.
Here, it is integrated using the explicit Euler method 
\begin{equation}
    \begin{aligned}
        \hat{H}_{k+1} &= \hat{H}_k + \tau \frac{\mathrm{d}\hat{H}}{\mathrm{d}t} \\
                      &= \hat{H}_k + \Delta \hat{H}_{\bm{\theta}}(\hat{H}_k, B_k, B_{k+1}),
    \end{aligned}
    \label{eq:ja_forward}
\end{equation}
where $\frac{\mathrm{d}\hat{H}}{\mathrm{d}t}$ is approximated using~\eqref{eq:JA_dmdh} and~\eqref{eq:JA_dhdt}. 

During training of this model (and following \gls{ja} variants), the physical parameters presented in~\eqref{eq:JA_parameters} are not directly tuned.
Instead, the tunable parameters are wrapped in the sigmoid function $\sigma$ and scaled with individual factors to produce the physical parameters.
This results in the tunable parameters
\begin{equation}
    \bm{\theta} = \sigma^{-1}\left(
    \begin{bmatrix}
        \frac{M_\mathrm{s}}{\eta_{M_\mathrm{s}}}, \frac{a}{\eta_a}, \frac{\alpha_\mathrm{W}}{\eta_{\alpha_\mathrm{W}}}, \frac{k^\prime}{\eta_{k^\prime}}, \frac{c}{\eta_c}
    \end{bmatrix}^\top\right).
    \label{eq:JA_parameters}
\end{equation}
where $\eta_z \in \mathbb{R}_+$ $\forall z \in \{M_\mathrm{s}, a, \alpha_\mathrm{W}, k^\prime, c\}$ are heuristically chosen scaling factors, and $\sigma^{-1}$ is the inverse function of the sigmoid function that is applied element-wise to the contents of the input vector. This mapping is done to ensure that the parameters stay positive and roughly within their physical value ranges during numerical optimization.

Note that this basic \gls{ja} model is unable to properly discern between different operating points (e.g., changes in material behavior due to temperature or frequency), and provides the parameters that best describe all conditions on average.
This circumstance is addressed in the following model variant.

\subsubsection{GRU directly parameterizes JA (GRU-JADP)}
\label{subsubsec:gru_directly_parameterizes_ja}
In this setup, a \gls{gru} provides the parameters to the \gls{ja} model
\begin{equation}
    \begin{aligned}
        \bm{\theta}_\mathrm{JA} &= \bm{g}_k^{(:5)} = \mathrm{GRU}_{\bm{\theta}_\mathrm{GRU}}(\bm{x}_k, \bm{g}_{k-1})^{(:5)} \\
        \hat{H}_{k+1} &= \hat{H}_k + \Delta \hat{H}_{\bm{\theta}_\mathrm{JA}}(\hat{H}_k, B_k, B_{k+1}).
    \end{aligned}
    \label{eq:gru_feeding_ja}
\end{equation}
The first five elements of $\bm{g}_k^{(:5)}$ are interpreted as the parameters of the \gls{ja} model, while the remaining elements $\bm{g}_k^{(5:)}$ may carry further information.
The idea is that the operating-point dependence of the parameters is learned by the \gls{gru}.

\subsubsection{JA and residual GRU}
\label{subsubsec:ja_and_residual_gru}
This version aims to use the basic behavior of the \gls{ja} model and correct the prediction with a \gls{gru} cell.
This is implemented by summing the outputs of~\eqref{eq:ja_forward} and~\eqref{eq:direct_prediction_function}
\begin{equation}
    \hat{H}_{k+1} = \underbrace{\hat{H}_k + \Delta \hat{H}_{\bm{\theta}_\mathrm{JA}}(\hat{H}_k, B_k, B_{k+1})}_{\mathrm{JA}} + \underbrace{\Delta \hat{H}_{\bm{\theta}_\mathrm{GRU}}\left(\bm{x}_k, \bm{g}_{k-1} \right)}_{\mathrm{GRU}}.
    \label{eq:ja_and_residual_gru}
\end{equation}
This model has multiple issues, which is why it is omitted from the final comparison.
The model training is fairly unstable when both models are trained together and requires double precision floating point computations, heavily increasing the time required for training the model. 
Additionally, it performed worse than the setups without the \gls{ja} model.
It seems that the information provided by the \gls{ja} part is negligible or even a hindrance for the \gls{gru} reinforcing the theoretical objections.

\subsubsection{Physics-informed neural network with JA regularization}
\label{subsubsec:pinn_ja}
Here, the prediction of a \gls{ja} model is used to add a regularization for the training loss function of the \gls{gru} inspired by physics-informed machine learning and \gls{pinn} (see, e.g.,~\cite{Karniadakis2021, Meng2025}), which is an intensively researched topic where prior knowledge in the form of physical equations is incorporated into data-driven models. The underlying idea is that the physical equation supports the training process in terms of convergence speed and robustness against measurement noise. Furthermore, it should provide a better extrapolation accuracy compared to a purely data-driven model.

The prediction function for this model type is the same as in~\eqref{eq:direct_prediction_function}%
\begin{equation}
        \hat{\tilde{H}}_k = \hat{\tilde{H}}_{\bm{\theta}_\mathrm{GRU}}\left(\bm{x}_k, \bm{g}_{k-1} \right) = \bm{g}_{k}^{(0)},
    \label{eq:pinn_ja}
\end{equation}
but the training loss is 
\begin{equation}
    \mathcal{L}_{\mathrm{JA}, \mathrm{GRU}} = \mathcal{L}^\prime_\mathrm{\change{a}RMSE} + \lambda_w \mathcal{L}_\mathrm{JA},
    \label{eq:pinn_training_loss}
\end{equation}
where the regularization term $\mathcal{L}_\mathrm{JA}$ is
\begin{equation}
        \mathcal{L}_\mathrm{JA} = \sqrt{\frac{1}{k_2-k_1 +1} \sum_{k=k_1}^{k_2}\left| e_\mathrm{JA} \right|^2},
    \label{eq:loss_JA}
\end{equation}
with 
\begin{equation}
    e_\mathrm{JA} = \Delta \hat{H}_{\bm{\theta}_\mathrm{JA}}(\hat{H}_{k-1}, B_{k-1}, B_{k}) - (\hat{H}_{k} - \hat{H}_{k-1}).
\end{equation}
The weighting factor between the two loss components is $\lambda_w \in \mathbb{R}_+$ .
Note that $\bm{\theta}_\mathrm{JA}$ and $\bm{\theta}_\mathrm{GRU}$ are trained at the same time with this specific setup.

In the presented setup, the potential of the PINN approach is likely not properly utilized. 
This approach should be providing the \gls{gru-p} with an extrapolation railing, but the \gls{ja} is likely incapable of providing significant help due to its relatively poor prediction performance.

\subsection{Preisach}
\label{subsec:preisach}

The Preisach model~\cite{Preisach1935} is a phenomenological approximation of the hysteresis effect observed in magnetic materials. 
A differentiable version of the Preisach model akin to the one presented in~\cite{Roussel2022} has been implemented\footnote{Open source repository corresponding to~\cite{Roussel2022} for the differentiable Preisach model: \url{https://github.com/roussel-ryan/diff_hysteresis}.}. Due to its differentiability, the model may be optimized via gradient descent in the same framework as the other models.

The main components of the Preisach model are the hysteron density function \mbox{$\mu(\alpha, \beta): \mathbb{R} \times \mathbb{R} \rightarrow \mathbb{R}$} and the hysteron operator \mbox{$\gamma_{\alpha, \beta}$}.
The prediction function of the differentiable Preisach model is~\cite{Roussel2022}
\begin{equation}
    B_k = \sum_{i=0}^{N-1} \bm{\mu}_i \gamma(H_k, H_{k-1}, \bm{\gamma}_{k-1, i}, \alpha_i, \beta_i, T) = \sum_{i=0}^{N-1} \bm{\mu}_i \bm{\gamma}_{k, i},
    \label{eq:preisach_pred_function}
\end{equation}
where $\bm{\mu}_i = \mu(\alpha_i, \beta_i)$ is a discretized approximation of the hysteron density function with \mbox{$i \in \{0, \dots N-1\}$}, and
\begin{equation}
\begin{aligned}
    & \gamma(H_k, H_{k-1}, \gamma_{k-1}, \alpha_i, \beta_i, T) \\ 
    & =  \begin{cases}
      \mathrm{min}\left(\gamma_{k-1} + \mathrm{tanh}\left(\frac{H_k - \beta_i}{\|T\|}\right) , 1 \right) & \text{if $H_k > H_{k-1}$}\\
      \mathrm{max}\left(\gamma_{k-1} - \mathrm{tanh}\left(\frac{-H_k + \alpha_i}{\|T\|}\right), -1 \right) & \text{if $H_k \leq H_{k-1}$}\\
    \end{cases}
\end{aligned}
\end{equation}
is the differentiable hysteron operator (proposed in the supplementary material of~\cite{Roussel2022}), $T \in \mathbb{R}_+$ is a parameter controlling the rate of change for the hysteron operator ($T=10^{-3}$ was utilized for all computations), and $\bm{\gamma}_{k-1} \in \mathbb{R}^N$ are the values that were computed by the hysteron operators in the last step.

The model was implemented with a static, equally spaced grid of $\alpha_i$ and $\beta_i$ values and the corresponding values for the hysteron density vector $\bm{\mu} \in \mathbb{R}^N$ were optimized using trajectory rollout and stochastic gradient descent (see~\cite{Roussel2022}).
Furthermore, the output function~\eqref{eq:preisach_pred_function} was adapted to
\begin{equation}
    \hat{\tilde{B}}_k = \omega_2 \cdot \left( \sum_{i=0}^{N-1} \bm{\mu}_i \bm{\gamma}_{k, i} \right) + \omega_1 \tilde{H}_k + \omega_0,
    \label{eq:adapted_preisach_prediction_function}
\end{equation}
resulting in the following vector of tunable parameters for the differentiable Preisach model:
\begin{equation}
    \bm{\theta} = \begin{bmatrix}
        \omega_0 & \omega_1 & \omega_2 & \bm{\mu}
    \end{bmatrix}.
\end{equation}

\begin{figure}[htbp]
    \centering
    \includegraphics[width=\columnwidth]{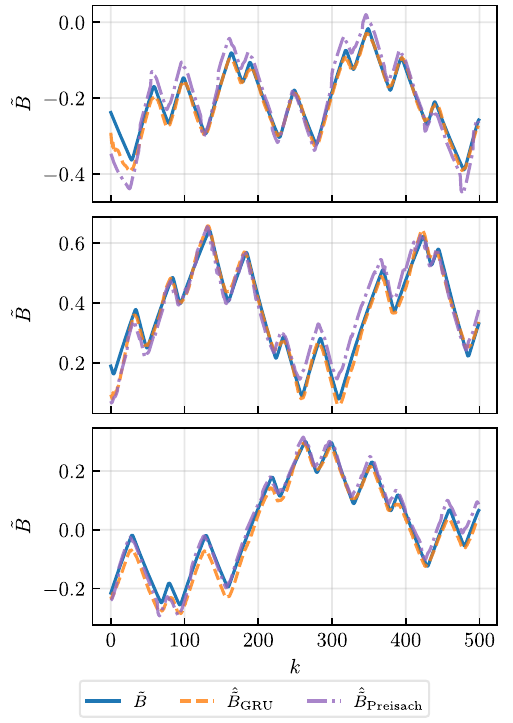}%
    \caption{Exemplary prediction performance for the \gls{gru-p} and the presented Preisach model variation. Note that within this preliminary evaluation, the normalized magnetic flux density $\tilde{B}$ was the prediction target and the normalized magnetic field $\tilde{H}$ was provided, which is reversed compared to the direction generally targeted within this work and \gls{mc2}.}
    \label{fig:qual_comparison_preisach}
\end{figure}

In a preliminary test, the Preisach model~\eqref{eq:adapted_preisach_prediction_function} was compared to a basic \gls{gru-p}.
Both were trained on the same training data and evaluated on the same unseen test data from the material 3C90 that was gathered with \mbox{$f_\mathrm{sw} = \SI{320}{\kilo\hertz}$} at \mbox{$\vartheta = \SI{25}{\celsius}$}.
Exemplary trajectories from the training data are depicted in Fig.~\ref{fig:qual_comparison_preisach} 
and multiple metrics for the prediction performance are reported in Tab.~\ref{tab:preliminary_quan_results_preisach}.
The considered performance metrics are the \gls{mse}
\begin{equation}
    \change{\mathcal{L}}_\mathrm{MSE} = \frac{1}{k_2 - k_1 + 1}\sum_{k=k_1}^{k_2} e_k^2,
    \label{eq:mse}
\end{equation}
with $e_{k} = \hat{\tilde{H}}_{k} - \tilde{H}_{k}$,
the \gls{mae}
\begin{equation}
    \change{\mathcal{L}}_\mathrm{MAE} = \frac{1}{k_2 - k_1 + 1}\sum_{k=k_1}^{k_2} |e_k|,
\end{equation}
and the \gls{wce}
\begin{equation}
    \change{\mathcal{L}}_\mathrm{WCE} = \max\left(\left\{
        |e_{k_1}|, |e_{k_1 + 1}|, \dots , |e_{k_2}|
    \right\}\right).
\end{equation}

\begin{table}[t]
    \caption{Initial test results \gls{gru-p} vs. Preisach}
    \centering
    \addtolength{\tabcolsep}{-0.1em}
    \begin{tabular}{lrr}\toprule
    Model &  Preisach & \gls{gru-p}\\
    \midrule
    \# Model params. & $328$ & $248$ \\
    Avg. $\change{\mathcal{L}}_\mathrm{MSE}$ & $0.0081$ & $0.0005$ \\
    Avg. $\change{\mathcal{L}}_\mathrm{MAE}$ & $0.0534$ & $0.0143$\\
    Avg. $\change{\mathcal{L}}_\mathrm{WCE}$ & $0.6763$ & $0.2417$\\
    \bottomrule
    \end{tabular}
    \label{tab:preliminary_quan_results_preisach}
\end{table}

Based on these initial tests, it was deemed that the model in the presented form could likely not come close to the parameter to accuracy ratio of the \gls{rnn}-based models.
Therefore, the inversion (prediction of $H$ based on $B$) and further optimizations were not investigated further.

\subsection{LLG-based modeling}
\label{subsec:llg}

The idea of this approach is once more to utilize physics-informed machine learning (cf. Sec.~\ref{subsubsec:pinn_ja}) to incorporate a physical equation with the hidden state of the \glspl{rnn}.
This could provide a physical interpretation for the hidden state, which could yield insights into the internal structure of the material, and give the \gls{rnn} a guideline on how the hidden state should propagate.

The \gls{llg} equation deals with the transfer of spin torques on a micromagnetic scale~\cite{Abert2025}:
\begin{equation}
    \frac{\partial \bm{m}}{\partial t} = - \frac{\gamma}{1 + \alpha_\mathrm{d}^2} \bm{m} \times \bm{H}_\mathrm{eff} - \frac{\alpha_\mathrm{d} \gamma}{1+\alpha_\mathrm{d}^2} \bm{m} \times \left(\bm{m} \times \bm{H}_\mathrm{eff}\right),
    \label{eq:LLG}
\end{equation}
where $\alpha_\mathrm{d}$ is a dimensionless damping parameter, $\gamma$ is the reduced gyromagnetic ratio, $\bm{m}$ is the unit-vector field representation of the magnetization, and $\bm{H}_\mathrm{eff}$ is the effective magnetic field resulting from all relevant interactions within the material~\cite{Abert2025}.
The effective field depends on the location of the magnetization in the material (and local magnetization components), therefore,~\eqref{eq:LLG} is a \gls{pde} governed via
\begin{equation}
    \frac{\partial \bm{m}(\bm{z})}{\partial t} = f\left(\bm{m}(\bm{z}), \bm{H}_\mathrm{eff}(\bm{z})\right),
    \label{eq:LLG_shorthand_PDE}
\end{equation}
which may be approximated on a discretized space (exemplary in 2D)
\begin{equation}
    \frac{\partial \bm{m}(i,j)}{\partial t} = f\left(\bm{m}(i,j), \bm{H}_\mathrm{eff}(i,j)\right),
    \label{eq:LLG_shorthand_PDE_discretized}
\end{equation}
where $(i, j)$ corresponds to the coordinates $\bm{z} = \Delta z \cdot \begin{bmatrix}i & j\end{bmatrix}^\mathrm{T}$.
The discretized \gls{pde} may be solved using typical numerical methods such as \gls{ode} solvers.

\subsubsection{LLG PDE simulation}
\label{subsubsec:llg_pde_simulation}
As a preliminary assessment, it was tested how well a simulation of the \gls{llg} equation may approximate macroscopic hysteresis effects by summing over the microscopic magnetization. For this, the Python micromagnetic simulation NeuralMag\footnote{Open-source repository for the utilized micromagnetic simulator: \mbox{\url{https://gitlab.com/neuralmag/neuralmag}}}~\cite{Abert2025} was utilized.

The main issue is that neither the spatial nor the temporal scale match the \gls{mc2} setting.
The data was acquired with \mbox{$\tau = \SI{62.5}{\nano\second}$}, while such micromagnetic simulations are generally performed at $\tau \approx \SI{10}{\pico\second}$ or with even smaller time steps (Fig. $10$ in~\cite{Vansteenkiste2014}), which would require roughly $6000$ steps per data point from the \gls{mc2} data base.
This makes the simulation at the required timescale computationally \change{highly challenging or even practically} infeasible.

\subsubsection{Vector field GRU (GRU-V)}
\label{subsubsec:vector_field_gru}
\begin{figure}[htbp]
    \centering
    \includegraphics[width=\columnwidth]{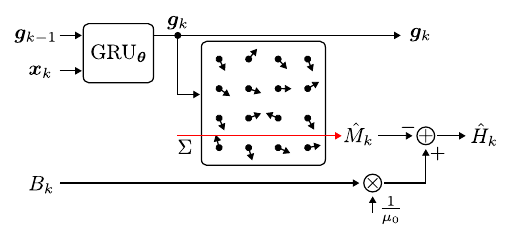}%
    \caption{One forward propagation step for the vector field \gls{gru}.}
    \label{fig:vector_fieldGRU}
\end{figure}

The hidden state $\bm{g}_k \in \mathbb{R}^{d_\mathrm{g}}$ for this model is arranged in a 2D grid of 2D vectors which are intended to approximate a discretized vector field of the magnetization, i.e., $\bm{g}_k(i,j) \in \mathbb{R}^{2}$ $\forall$ $i \in \{0,1,\dots, N_\mathrm{g}\}$ and $j \in \{0,1,\dots, N_\mathrm{g}\}$ with $N_\mathrm{g} = \sqrt{d_\mathrm{g} / 2} - 1 \in \mathbb{N}_+$ (cf. Fig.~\ref{fig:vector_fieldGRU}).
The prediction function for this model was intended to be
\begin{equation}
    \hat{H}_k = \hat{H}_{\bm{\theta}}\left(\bm{x}_k, \bm{g}_{k-1} \right) = \frac{{B}_k}{\mu_0} - \hat{M}_k,
\label{eq:prediction_LLG_GRU}
\end{equation}
where
\begin{equation}
    \hat{M}_k = \sum_{i=0}^{N_\mathrm{g}} \left( \sum_{j=0}^{N_\mathrm{g}} \bm{g}_k(i,j)^{(0)} \right).
\label{eq:prediction_magnetization}
\end{equation}
Due to the numerical issues already encountered for the \gls{gru-m} in \eqref{eq:mag_gru_pred_phys}, the prediction function
\begin{equation}
    \hat{\tilde{H}}_k = \hat{\tilde{H}}_{\bm{\theta}}\left(\bm{x}_k, \bm{g}_{k-1} \right) = \tilde{B}_k - \sum_{i=0}^{N_\mathrm{g}} \left( \sum_{j=0}^{N_\mathrm{g}} \bm{g}_k(i,j)^{(0)} \right)
\label{eq:prediction_vector_field_GRU}
\end{equation}
was utilized instead. Here, $\bm{g}_k(i,j)^{(0)}$ refers to the first element of the two-dimensional vector at grid position $(i,j)$, which is interpreted as the magnetization direction that coincides with the measurement data.

Initially, it was planned to use the \gls{llg} as a regularization term for the hidden state of this model so that the states can interact with each other and, therefore, provide the state with physical interpretability.
However, due to the infeasibility of the isolated \gls{llg} simulation, the idea was not investigated further.
Instead, the \gls{gru-v} is essentially a different prediction function compared to \gls{gru-p}.
One further issue is that due to the complexity of the hidden state's interpretation, it is not feasible to warm it up the same way as the \gls{gru-p}.

\subsection{Model overview}
\label{subsec:model_overview}

Tab.~\ref{tab:model_overview} provides a qualitative comparison of the different model types presented in this section.
Some of the models are not part of the empirical evaluation presented in Sec.~\ref{sec:results} due to their poor performance in prior experiments or their computational infeasibility. Whether a model is considered for empirical evaluation is reflected in the column "Emp. eval." in Tab.~\ref{tab:model_overview}.
The reported model-size range in the numbers of parameters is not intended to show a model's range of feasibility, it simply aims to give a rough range of parameters where the model should typically be applied.

\begin{table*}[t]
    \caption{Qualitative overview over all presented model types}
    \centering
    \addtolength{\tabcolsep}{-0.1em}
    \begin{tabularx}{\linewidth}{lllcrL}\toprule
        Model name &  Abbreviation & Section & Emp. eval. & Typ. \# params. & Descriptions \\
        \midrule
        GRU with direct prediction & GRU-P & \ref{subsubsec:GRU_direct_pred} & \cmark & $\left[50, 10000\right]$ &
        \begin{minipage}[t]{\linewidth}
        \begin{itemize}[leftmargin=*, topsep=0pt, itemsep=1pt]
            \item A \gls{gru}-based model, where the first element of the hidden state is used as the prediction for the normalized magnetic field
        \end{itemize}
        \end{minipage}
        \vspace{2pt}\\
        Magnetization GRU & GRU-M & \ref{subsubsec:magnetization_gru} & \cmark & $\left[50, 10000\right]$ & 
        \begin{minipage}[t]{\linewidth}
        \begin{itemize}[leftmargin=*, topsep=0pt, itemsep=1pt]
            \item A \gls{gru}-based model, where the first element of the hidden state is used as an approximation of the normalized magnetization
            \item The magnetic field prediction is then calculated using \eqref{eq:mag_gru_pred}
        \end{itemize}
        \end{minipage}
        \vspace{2pt}\\
        GRU parameterizes linear model & GRU-L & \ref{subsubsec:GRU_linear_model}  & \cmark & $\left[100, 10000\right]$ & 
        \begin{minipage}[t]{\linewidth}
        \begin{itemize}[leftmargin=*, topsep=0pt, itemsep=1pt]
            \item A \gls{gru}-based model, where the first element of the hidden state is used as the prediction for the normalized inverse of the permeability
        \end{itemize}
        \end{minipage}
        \vspace{2pt}\\
        LSTM with direct prediction & LSTM-P & \ref{subsec:lstm_with_direct_prediction} & \cmark & $\left[60, 10000\right]$ & 
        \begin{minipage}[t]{\linewidth}
        \begin{itemize}[leftmargin=*, topsep=0pt, itemsep=1pt]
            \item An \gls{lstm}-based model, where the first element of the hidden state is used as the prediction for the normalized magnetic field (analogous to \gls{gru-p})
        \end{itemize}
        \end{minipage}
        \vspace{2pt}\\
        Basic (inverse) JA & JA & \ref{subsubsec:basic_inverse_ja} & \cmark & 5 &         
        \begin{minipage}[t]{\linewidth}
        \begin{itemize}[leftmargin=*, topsep=0pt, itemsep=1pt]
            \item Inverse \gls{ja} \gls{ode} solved using explicit Euler (based on~\cite{Zirka2012, Zaman2016})
            \item Static parameters are fit for all operating conditions
        \end{itemize}
        \end{minipage}
        \vspace{2pt}\\
        GRU directly parameterizes JA & GRU-JADP & \ref{subsubsec:gru_directly_parameterizes_ja} & \cmark & $\left[50, 10000\right]$ &
        \begin{minipage}[t]{\linewidth}
        \begin{itemize}[leftmargin=*, topsep=0pt, itemsep=1pt]
            \item Inverse \gls{ja} \gls{ode} solved using explicit Euler (based on~\cite{Zirka2012, Zaman2016})
            \item The parameters are adapted dynamically using a \gls{gru}
        \end{itemize}
        \end{minipage}
        \vspace{2pt}\\
        JA and residual GRU & & \ref{subsubsec:ja_and_residual_gru} & \xmark & $\left[55, 10000\right]$ & 
        \begin{minipage}[t]{\linewidth}
        \begin{itemize}[leftmargin=*, topsep=0pt, itemsep=1pt]
            \item Sums the outputs of a \gls{gru} and \gls{ja}
        \end{itemize}
        \end{minipage}
        \vspace{2pt}\\
        \gls{pinn} with JA regularization & & \ref{subsubsec:pinn_ja} & \xmark & $\left[55, 10000\right]$ & 
        \begin{minipage}[t]{\linewidth}
        \begin{itemize}[leftmargin=*, topsep=0pt, itemsep=1pt]
            \item Uses a \gls{gru} analogously to \gls{gru-p} for the prediction
            \item A regularization term is added to the training loss comparing the \gls{gru} prediction with the \gls{ja} prediction
        \end{itemize}
        \end{minipage}
        \vspace{2pt}\\
        Preisach & & \ref{subsec:preisach} & \xmark & $\geq 300$ & 
        \begin{minipage}[t]{\linewidth}
        \begin{itemize}[leftmargin=*, topsep=0pt, itemsep=1pt]
            \item A differentiable Preisach model based on~\cite{Roussel2022}
        \end{itemize}
        \end{minipage}
        \vspace{2pt}\\
        \gls{llg} PDE simulation & & \ref{subsubsec:llg_pde_simulation} & \xmark & n.a. &
        \begin{minipage}[t]{\linewidth}
        \begin{itemize}[leftmargin=*, topsep=0pt, itemsep=1pt]
            \item A micromagnetic \gls{pde} based on~\cite{Abert2025}
        \end{itemize}
        \end{minipage}
        \vspace{2pt}\\
        Vector field GRU & GRU-V & \ref{subsubsec:vector_field_gru} & \cmark & $\left[1000, 10000\right]$ & 
        \begin{minipage}[t]{\linewidth}
        \begin{itemize}[leftmargin=*, topsep=0pt, itemsep=1pt]
            \item A \gls{gru}-based model, where the hidden state is interpreted as a 2D vector field of magnetization
            \item The prediction is performed by processing the magnetization estimates according to \eqref{eq:prediction_vector_field_GRU}
        \end{itemize}
        \end{minipage}
        \vspace{2pt}\\
        \bottomrule
    \end{tabularx}
    \label{tab:model_overview}
\end{table*}

\section{Results}
\label{sec:results}

\change{The proposed models are implemented in Python mainly utilizing libraries from the JAX ecosystem~\cite{JaxEcosystem2020}, and the Equinox library~\cite{Kidger2021equinox}.}
The models are evaluated on a subset of $5$ of the $15$ materials provided in the MagNetX database, each of which holds a specific challenge for modeling.
An overview of the $5$ testing materials is provided in Tab.~\ref{tab:testing_materials}.
After the conclusion of \gls{mc2}, a training data set $\mathcal{D}^{(m_2)}_\mathrm{train}$ and a testing data set $\mathcal{D}^{(m_2)}_\mathrm{test}$ are available for each of the materials.
\change{A single model is always trained on  $\mathcal{D}^{(m_2)}_\mathrm{train}$ and tested on $\mathcal{D}^{(m_2)}_\mathrm{test}$ for the same material $m_2$. 
The split between training and testing data sets as utilized in this work has been decided on by the MC2 organizing committee.
No cross-material modeling is evaluated within the scope of this work.}
\change{It should also be noted that the "unseen test data" is not necessarily extrapolation data. It generally comprises data from the training distribution that was held back, or it is data that requires extensive interpolation, e.g., the missing frequencies for material 3C95 (see Tab.~\ref{tab:testing_materials}).}

\begin{table}[t]
    \caption{Material challenge overview}
    \centering
    \addtolength{\tabcolsep}{-0.1em}
    \begin{tabularx}{\linewidth}{ll}\toprule
    Material name & Challenge \\
    \midrule
    3C92 & Small $\mathcal{D}^{(\mathrm{A})}_\mathrm{train}$ and switching artifacts \\
    3C95   & Missing $f_\mathrm{sw} = \{80, 500\} \, \SI{}{\kilo\hertz}$ in $\mathcal{D}^{(\mathrm{B})}_\mathrm{train}$\textbf{}\\
    FEC007 & Only one step for warmup in $\mathcal{D}^{(\mathrm{C})}_\mathrm{test}$\\
    FEC014 & Missing B-values in $\mathcal{D}^{(\mathrm{D})}_\mathrm{train}$ \\
    T37    & Filter material with high permeability \\
    \bottomrule
    \end{tabularx}
    \label{tab:testing_materials}
\end{table}

\subsection{Model selection for submission to MC2}

\begin{table}[t]
    \caption{Submitted model performance overview}
    \centering
    \addtolength{\tabcolsep}{-0.1em}
    \begin{tabularx}{\linewidth}{lrrrrr}\toprule
    & 3C92 & 3C95 & FEC007 & FEC014 & T37\\
    \midrule
    \# Train seq. in $\mathcal{D}^{(m_2)}_\mathrm{train}$    & 105 & 6,758 & 8,158 & 9,536 & 6,901 \\
    \# Test seq. in $\mathcal{D}^{(m_2)}_\mathrm{test}$   & 1,260 & 945 & 1,116 & 1,242 & 1,080 \\
    \# Model params.    & $325$ & $325$ & $325$ & $325$ & $325$ \\
    Model size   & \SI{3}{\kilo\byte}  & \SI{3}{\kilo\byte} & \SI{3}{\kilo\byte} & \SI{3}{\kilo\byte} & \SI{3}{\kilo\byte} \\
    \midrule
    Avg. SRE (in $\%$)         & $10.98$ & $6.51$ & $9.40$ & $6.04$ & $7.16$ \\
    95th-per. SRE (in $\%$)    & $22.99$ & $17.46$ & $27.63$ & $16.20$ & $18.38$ \\
    Avg. NERE (in $\%$)        & $0.65$ & $1.23$ & $1.82$ & $1.09$ & $0.56$ \\
    95th-per. NERE (in $\%$)   & $2.03$ & $3.96$ & $4.91$ & $4.18$ & $1.73$ \\
    \bottomrule
    \end{tabularx}
    \label{tab:finalResults}
\end{table}

\begin{figure*}[htbp]
    \centering
    \includegraphics[]{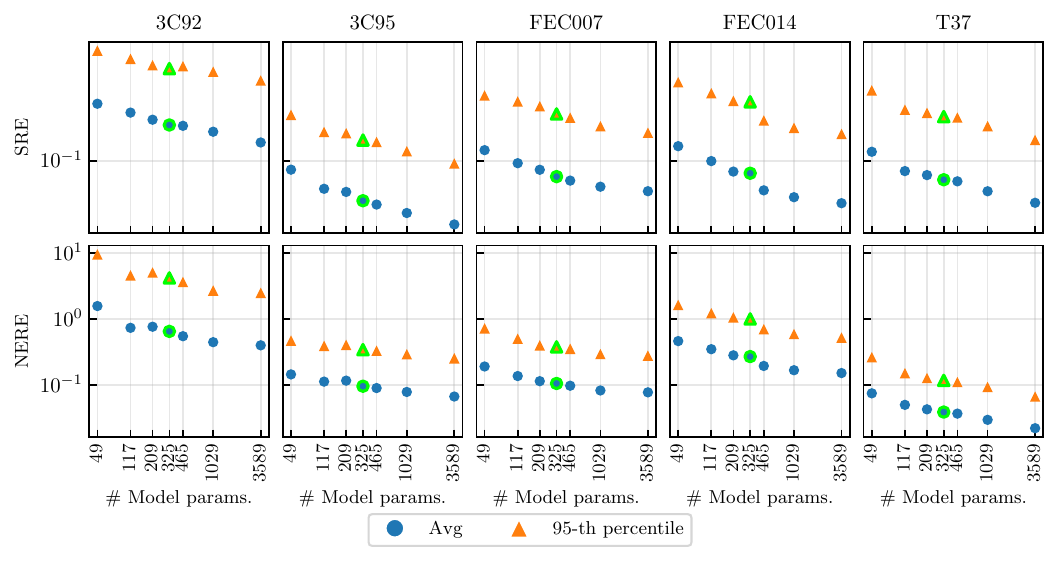}
    \caption{A Pareto front showing the accuracy to model size tradeoff is depicted for the submitted model type \gls{gru-p}. The error values are averaged over multiple trials to reduce the effect of particularly good or bad initialization of the model. The depicted values are computed using cross validation on the training data, as the final testing data was not yet available. The model setup highlighted in green has been submitted to \gls{mc2} and has been used for the results shown in Tab.~\ref{tab:finalResults}, Fig.~\ref{fig:BH_curve_B_800}, and Fig.~\ref{fig:BH_curve_D_125}.}
    \label{fig:pareto_front_submission}
\end{figure*}

\begin{table}[t]
    \caption{\change{Training parameters for the submitted models}}
    \centering
    \addtolength{\tabcolsep}{-0.1em}
    \begin{tabularx}{\linewidth}{>{\changetablecolor}l >{\changetablecolor}r >{\changetablecolor}l}\toprule
    Parameter & Value & Comment\\
    \midrule
    Optimizer & Adam & Optax Python implementation~\cite{JaxEcosystem2020} \\
    Init. learning rate & $10^{-3}$ & \begin{minipage}[t]{\linewidth} Scheduled to decay exponentially,\\ starts after $2000$ optimization steps \end{minipage} \\
    Final learning rate & $10^{-4}$ & \\
    Decay rate & $0.1$ & \\
    \# Epochs & $1500$ & \begin{minipage}[t]{\linewidth} Adapted to $10000$ for 3C92 and\\ $2500$ for T37, no early stopping \end{minipage} \\
    Batch size $b$ & $512$ & \\
    Subsequence len. $l$ & $156$ & Adapted to $129$ for FEC007\\
    Past length & $28$ & Adapted to $1$ for FEC007\\
    Future length & $128$ & \\
    \# Seeds & $5$ & \\
    \bottomrule
    \end{tabularx}
    \label{tab:training_parameters}
\end{table}

The final model selection for the submission to \gls{mc2} was done before $\mathcal{D}^{(m_2)}_\mathrm{test}$ was available in order to have a true training-test split.
Therefore, the following considerations are based solely on $\mathcal{D}^{(m_2)}_\mathrm{train}$ for each of the materials.

From previous tests, the decision was already made to utilize the \gls{gru-p} model. 
An important aspect of this model is the size, which is defined by the number of hidden units and the number of inputs. 
In the following, the model size will be changed by varying the number of hidden units in the \glspl{rnn}.
The tradeoff between the model size measured by the number of parameters and the model accuracy is to be investigated.
The model accuracies are measured using the \gls{sre}~\eqref{eq:SRE} and \gls{nere}~\eqref{eq:NERE} scores on unseen test data.
For the following test, for each of the materials $\mathcal{D}^{(m_2)}_\mathrm{train}$ is split into a separate training $\mathcal{D}^{(m_2)}_\mathrm{train, train}$, evaluation $\mathcal{D}^{(m_2)}_\mathrm{train, eval}$, and test set $\mathcal{D}^{(m_2)}_\mathrm{train, test}$.

Overall, the choice for the submitted models was made to balance model size and accuracy motivated by the preliminary Pareto investigation depicted in Fig.~\ref{fig:pareto_front_submission}. 
However, it is not claimed that this is the best choice for the general case. For instance, if the chosen model application permits a larger computational budget and memory footprint or if there are specific accuracy constraints to the performance of the model, a larger or even smaller model might be better suited. 
Hence, the favorable scaling of accuracy with model size of the GRU-based approach through the choice of cell count is one of its strengths and allows for easy adaptation depending on application requirements.

\begin{figure}
    \centering
    \subfloat[Exemplary time series prediction.]{\includegraphics[width=0.98\columnwidth]{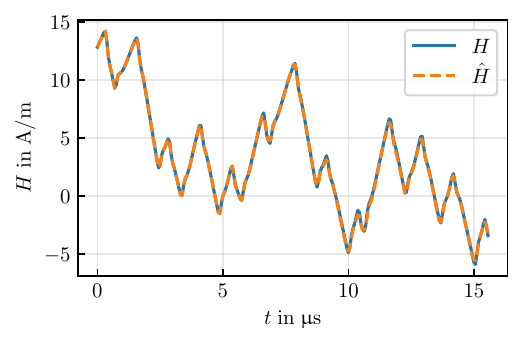}
    }%
    \\ 
    \subfloat[Exemplary $BH$-loop prediction for a full major loop.]{
        \includegraphics[width=0.98\columnwidth]{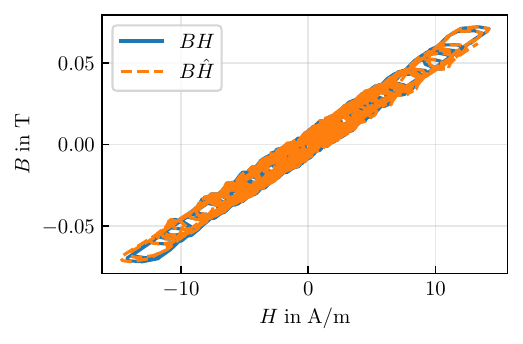}
    }%
    \caption{Performance of the \gls{gru-p} with $325$ parameters for the material 3C95 at $f_\mathrm{sw} = \SI{800}{\kilo\hertz}$ and $\vartheta = \SI{25}{\celsius}$.}
    \label{fig:BH_curve_B_800}
\end{figure}

\begin{figure}
    \centering
    \subfloat[Exemplary time series prediction.]{\includegraphics[width=0.98\columnwidth]{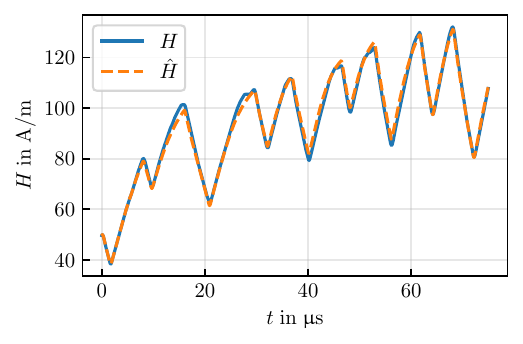}
    }%
    \\ 
    \subfloat[Exemplary $BH$-loop prediction for a full major loop.]{
        \includegraphics[width=0.98\columnwidth]{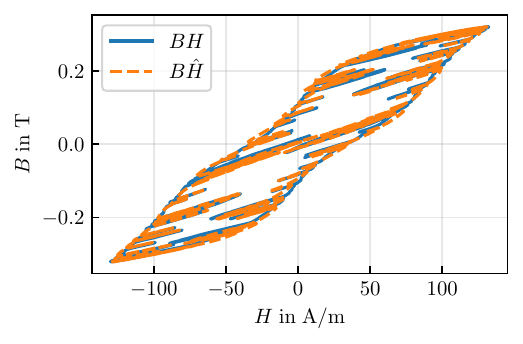}
    }%
    \caption{Performance of the \gls{gru-p} with $325$ parameters for material FEC014 at $f_\mathrm{sw} = \SI{125}{\kilo\hertz}$ and $\vartheta = \SI{25}{\celsius}$.}
    \label{fig:BH_curve_D_125}
\end{figure}

Tab.~\ref{tab:finalResults} summarizes key performance indicators of the $8$ cell GRU models that were submitted as final results for the \gls{mc2}. The provided model size is measured on disk (without any pruning or quantization).
\change{Tab.~\ref{tab:training_parameters} shows training parameters for the submitted models.}
In Fig.~\ref{fig:BH_curve_B_800} and Fig.~\ref{fig:BH_curve_D_125}, the prediction performance of the submitted model architecture is shown for representative full sequences from unseen data.

\subsection{\change{Ablation studies}}

\change{Two ablation studies are used to quantify the effectiveness of the proposed training loss formulation~\eqref{eq:weighted_rmse_loss} and warmup procedure.
For the training loss ablation study, the results for four different training losses are compared:
The "\gls{mse}"-case utilizes~\eqref{eq:mse}, while
the "adapted \gls{rms}"-case utilizes~\eqref{eq:weighted_rmse_loss} as the training loss.
The "adapted RMS (only $H^{-1}_\mathrm{RMS}$)"-case uses
\begin{equation}
    \mathcal{L}_{\mathrm{oH}} = \sqrt{\mathcal{L}_\mathrm{MSE}} \cdot H_{\mathrm{max}} \cdot  \left( \frac{1}{k_3 + 1} \sum_{k=0}^{k_3} H_k^2 \right)^{-\frac{1}{2}},
\end{equation}
while the "adapted RMS (only $\Delta B$)" case uses \eqref{eq:rmse_loss} directly as the training loss without the normalization with the sequence \gls{rms}.
}

\begin{figure*}[htbp]
    \centering
    \includegraphics[width=\linewidth]{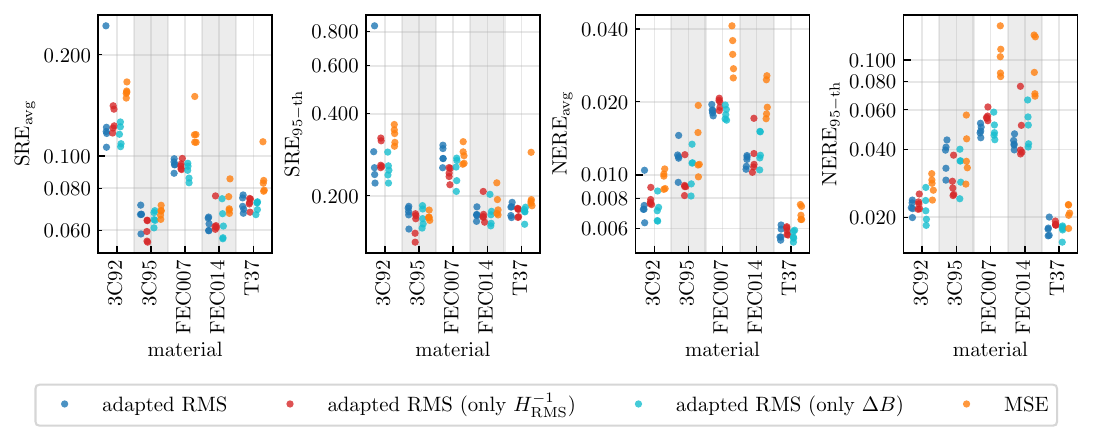}
    \caption{\change{An ablation study is performed to assess the effect of different training loss formulations. The figure shows the average and $95$-th percentile \gls{sre} and \gls{nere} scores for $5$ models per loss setup per material.}}
    \label{fig:ablation_loss}
\end{figure*}

\change{Fig.~\ref{fig:ablation_loss} shows quantitative results for the ablation loss study.
Both adaptations (division by $H_\mathrm{RMS}$ and multiplication by $\Delta B$) result in an improvement compared to the \gls{mse}. However, the combination of the two in the form of the adapted RMS is not clearly better or worse than either isolated version.
This is likely because both adaptations have the same effect of reducing emphasis on the strongly saturated cases that are difficult for the model to approximate. 
}

\change{
For the warmup procedure ablation study, four setups are compared as well:
The default setup denoted as "replace H (seq - seq)" uses the full available $H_{k_0:k_1 -1}$ sequence for warming up according to~\eqref{eq:H_past_injection} in both training and inference on the test set. The setup denoted as "replace H (one - seq)" does not utilize~\eqref{eq:H_past_injection} during training but only initializes the state before the prediction starts as \begin{align}
    \bm{g}^\prime_{k_1 - 1} = \begin{bmatrix}\tilde{H}_{k_1 - 1} \\ \bm{0}\end{bmatrix}.
    \label{eq:shortened_warmup}
\end{align}
During inference, it utilizes~\eqref{eq:H_past_injection} identically to the default setup.
The third setup "replace H (one - one)" utilizes~\eqref{eq:shortened_warmup} in both the training and the inference and, therefore, only utilizes the last element of the warmup sequence for inference. Lastly, the "init zeros" setup initializes the state to zero at $k_0$ and lets it develop from there. Note that for this setup $H_{k_0:k_1 -1}$ is not utilized as there is no mechanism available to feed it to the \gls{gru} cell.
For material FEC007, there is no warmup sequence provided and the three setups that utilize known $H$ values are essentially identical.}

\begin{figure*}[htbp]
    \centering
    \includegraphics[width=\linewidth]{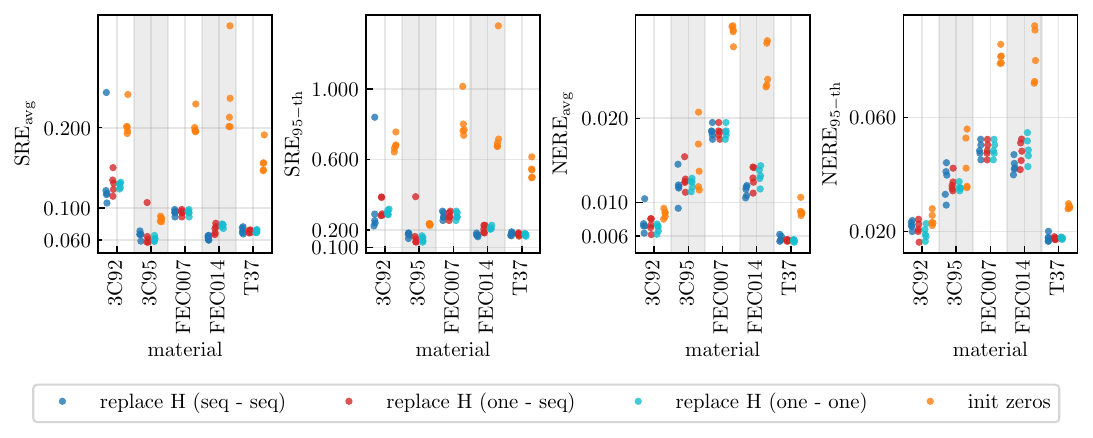}
    \caption{\change{An ablation study is performed to assess the effect of different intialization strategies. The figure shows the average and $95$-th percentile \gls{sre} and \gls{nere} scores for $5$ models per warmup approach per material.}}
    \label{fig:ablation_init}
\end{figure*}

\change{
Fig.~\ref{fig:ablation_init} shows quantitative results for the ablation warmup study.
The results show that initializing the hidden state to zeros produces poor prediction results. 
This is to be expected, as the model lacks sufficient contextual information about the current position in the hysteresis loop.
From the comparison between the other three setups, it can be seen that using the full sequence both in training and in inference does not have a significant impact on most of the results.
This means that injecting H into the state during warmup does not introduce significant training instability.
However, continuously injecting a whole sequence of values also does not provide significant improvements in model performance.
A slight trend can be seen for material FEC014, where the training and inference with the full sequences improve on the two other cases with single-step training and single-step inference warmup. A reason for this could be that FEC014 has the strongest hysteresis behavior and, therefore, more context may be required to initialize the model and identify the hysteretic state accurately.
}

\subsection{\change{Wall-clock time}}
\label{subsec:wall_time_exps}
\begin{figure}[htbp]
    \centering
    \includegraphics[width=\linewidth]{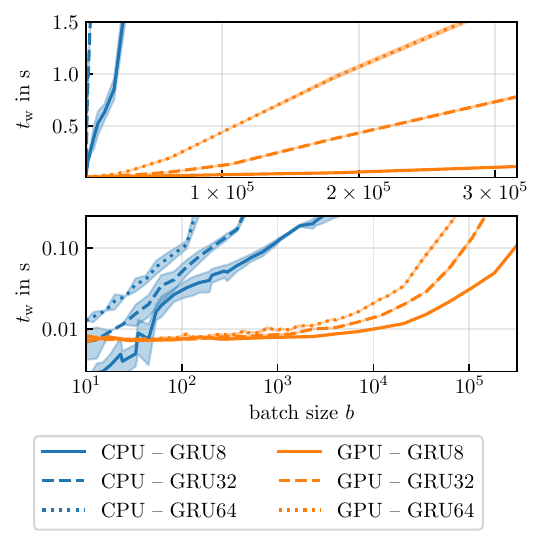}
    \caption{\change{Wall-clock times for evaluating a sequence of $100$ warmup and $900$ model prediction steps on GPU and CPU with $b$ in parallel. The two subfigures show the same data, once in linear scale, once in double logarithmic scale. The lines show the average computation time of $100$ runs per model and batch size, while the shaded area represents the variance.}}
    \label{fig:wall_time}
\end{figure}

\change{To provide context about the computational complexity of the model inference, wall-clock time tests are performed on a workstation with an AMD Ryzen Threadripper PRO 7955WX CPU with $16$ cores and \SI{4.5}{\giga\hertz} clock frequency, $128$ GB of DDR5 RAM, and an NVIDIA RTX 6000 Ada generation GPU. 
Three model sizes (GRU8 with $325$, GRU32 with $3589$, and GRU64 with $13317$ parameters) are used to predict a sequence with $100$ warmup and $900$ prediction steps.
This is done with a batch size $b$ of $10$ to approx.~$3.16 \times 10^5$ to emulate parallel processing (e.g., for numerical optimization or \gls{fem} solving, both of which benefit greatly from massive parallelization). Each combination of $b$ and the model type is repeated 100 times to average out random contributions.
The results are presented in Fig.~\ref{fig:wall_time}.}

\change{
The results show that the GPU enables massive parallelization in contrast to the CPU.
However, for very small batch and model sizes, the CPU implementation is faster.
Additionally, on the GPU, the smaller model with $325$ parameters allows a computation with batch size  $b = 3.16 \times 10^5$ at an average computation time of $\SI{0.110}{\second}$ compared to $\SI{1.722}{\second}$ for a model with $13317$ parameters, which constitutes a speedup by a factor of $\approx 15.65$ for the smaller model.
}

\subsection{Extended Pareto investigation}
\label{subsec:extended_pf}
\begin{figure*}[htbp]
    \centering
    \includegraphics[width=\textwidth]{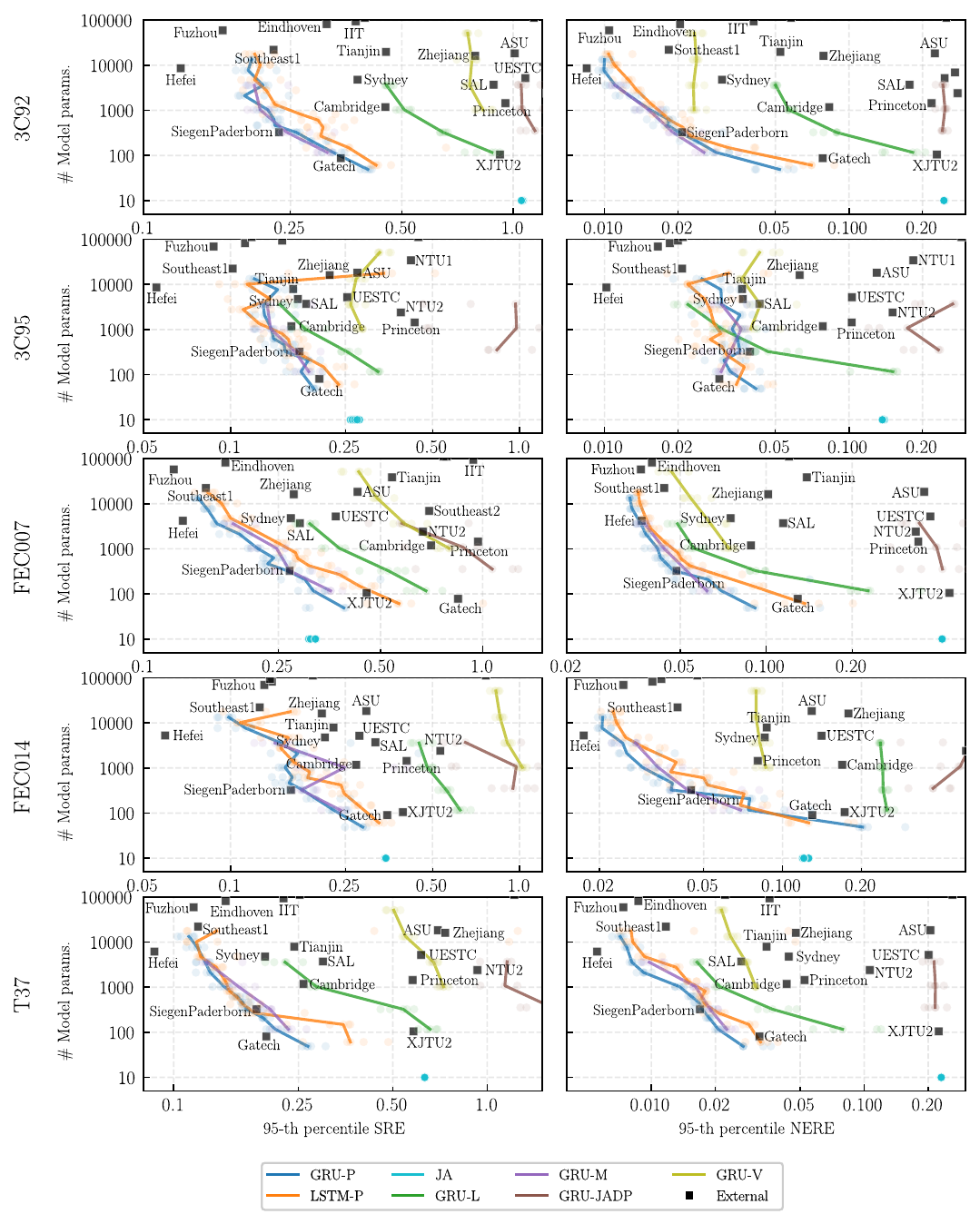}
    \caption{The accuracy to model size tradeoff for most of the model types introduced previously is depicted for the final $5$ materials. Individual trials are shown as lightly colored points. The median error of the trials for each model class is depicted as a line. Additionally, some other final submissions to the \gls{mc2} are shown as black squares. Note that the \gls{ja} model is depicted with $10$ parameters, but only requires them to fit into the same framework as the other models. In practice, it only requires $5$ of those parameters to function properly.}
    \label{fig:pareto_front}
\end{figure*}

With $\mathcal{D}^{(m_2)}_\mathrm{test}$ available, a further Pareto investigation is performed to evaluate the previous model selection based only on $\mathcal{D}^{(m_2)}_\mathrm{train}$ and put the proposed model structures into relation with other submissions to \gls{mc2}.
For a variety of the previously introduced model types and \change{independently for each of the materials in $M_2$}, a collection of differently seeded trials is performed\change{, yielding one trained model per seed, model type, and material combination}. 
In Fig.~\ref{fig:pareto_front}, the results of this investigation are visualized.
The performance of each of the individual models is depicted as a lightly colored point, while the median error of each model type through the model sizes is shown as a line.
Additionally, the performance of models submitted for \gls{mc2} by other participating teams\footnote{The performance values for the external results can be found in the \gls{mc2} GitHub repository: \url{https://github.com/minjiechen/magnetchallenge-2}.} are shown.
Model performance is measured by the \gls{sre} and \gls{nere} scores on the unseen data in $\mathcal{D}^{(m_2)}_\mathrm{test}$.

The results indicate that the rather standard data-driven models are roughly within the same performance band, while the \gls{gru}-based models seem to be slightly more parameter efficient compared to the \gls{lstm-p} for most of the considered sizes.
Additionally, it seems that further structure provided by the \gls{ja} or, e.g., \gls{gru-l} and \gls{gru-v} deteriorates the performance.
Possible explanations are that the utilized structures do not align with the physical process and therefore do not provide any gains or it might be due to implementation issues. 
For instance, the \gls{gru-v} and \gls{gru-jadp} cannot be warmed up with the usually  employed process, as there is no suitable interpretation of the hidden state available.

Since at the time of writing, the model structures for the other submissions for \gls{mc2} are not openly available (as in open source), the model sizes cannot be  varied easily. \change{Hence, the results for the other teams as reported by the \gls{mc2} organizing committee are integrated into Fig.~\ref{fig:pareto_front} and no sweep through model sizes is performed for the other submitted models.}
Overall, the proposed \gls{gru-p} seems to perform very well for small model sizes, but does not seem to reach the performance of the model submitted by team 'Hefei'\change{~\cite{Wang2026}} for larger model sizes.
Whether this is due to the structure of the model or external factors such as parameterization or training structure will need to be investigated in the future.

It should be noted that the high accuracy at small model size could be especially interesting for incorporation into \gls{fem} simulation, where the material model must be evaluated for each node.
A smaller, albeit slightly less accurate, model might be preferable, as the difference between 300 and \change{10000} parameters results in significant differences in memory and computational effort when scaled with the number of nodes in the simulation mesh \change{as indicated in the wall-clock time experiments in Section~\ref{subsec:wall_time_exps}.}

\section{Conclusion}
\label{sec:conclusion}

A variety of modeling structures for the prediction of time-resolved magnetic material behavior \change{on a per material basis} were developed, implemented, and compared.
It was found that a largely data-driven model utilizing a single \gls{gru} layer could provide excellent parameter efficiency in terms of model size to model accuracy for small model sizes.
\change{In the scope of this work, further incorporation of physically motivated ideas could not improve the performance of the models.
Whether this is due to insufficient refinement in terms of the implementation of the models or due to the chosen model structures remains an open question at this point.
As a result,} it remains unclear how to properly inject physically motivated structure into the model and which physical equations are actually helpful for the macroscopic problem at hand.

\change{The reported accuracies are based on the data from the \gls{mc2} data set presented in Section~\ref{subsec:mc2_data_set} and are limited to the available switching frequency, temperature and flux density amplitude ranges and the considered switching patterns. Extrapolation beyond these boundaries is not considered and needs to be evaluated in future work.}
\change{The training and test data sets should also be extended with further material data, e.g., from powder cores and for core geometries other than toroidal cores to investigate whether the proposed model architectures are applicable to other material types and geometries.}
\change{To complement the work further}, a hyperparameter optimization for the proposed model structures should be performed in order to ensure that they perform as best as possible. 
Furthermore, the presented Pareto investigations should be extended with the other models that were submitted for \gls{mc2} to enable proper comparison throughout different parameterizations.
Additionally, future work should examine transfer learning for the provided model structures such that the model can be simply adapted to a new material with only limited measurement data (cf.~\cite{Li2023} for a steady-state example of this).

\section*{Acknowledgment}
The authors would like to thank the organizers and sponsors of
the \gls{mc2} for hosting a great, educational event
within the power electronics community. This article would not
have been possible without the provision of the MagNetX dataset
and the organizers’ tremendous effort administrating the challenge.

This research work was partially funded by the German Research Foundation (DFG) (grant ID: 467840481) and by the German Federal Ministry of Education and Research (BMBF) (grant ID: 01IS22081).

In preparing this manuscript, the authors made use of Claude (Sonnet 4.6), a large language model developed by Anthropic, to assist with the reformulation and linguistic refinement of selected passages. All scientific content, interpretations, and conclusions are solely the work of the authors, who bear full responsibility for the accuracy and integrity of this publication.

\bibliographystyle{IEEEtran}
\bibliography{bibliography}


\section{Biography Section}

\begin{IEEEbiography}[{\includegraphics[width=1in,height=1.25in,clip,keepaspectratio]{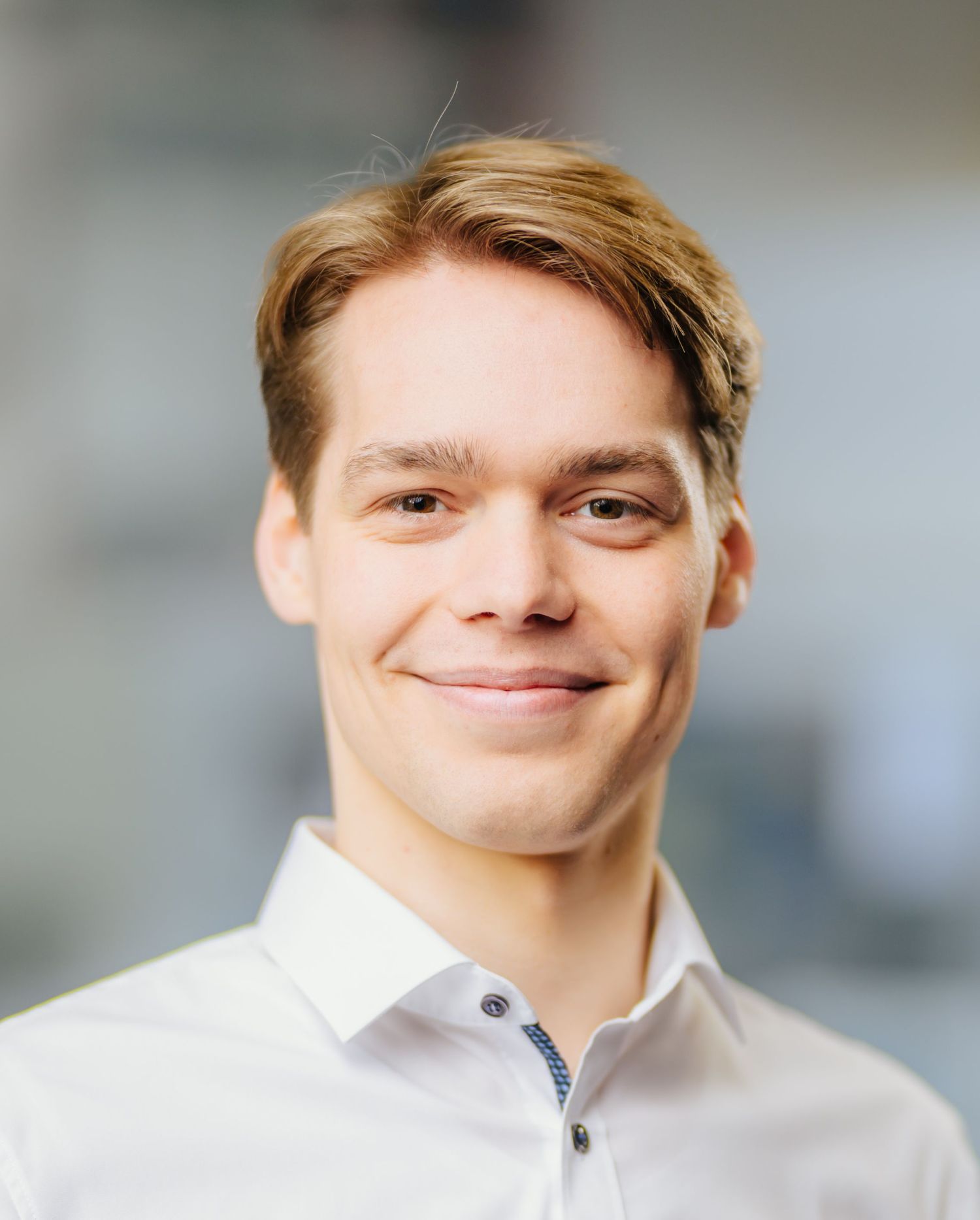}}]{Hendrik Vater} received the bachelor's and master's degree in electrical engineering from Paderborn University, Paderborn, Germany, in 2020 and 2023, respectively. 

He worked as a Research Associate with the Department of Power Electronics and Electrical Drives, Paderborn University, from 2023 until 2025.
Since 2026, he is working as a Research Associate at the Chair of Interconnected Automation Systems (IAS), University of Siegen, Siegen, Germany. His research interests include system excitation, system identification, and intelligent control methods.
\end{IEEEbiography}

\begin{IEEEbiography}[{\includegraphics[width=1in,height=1.25in,clip,keepaspectratio]{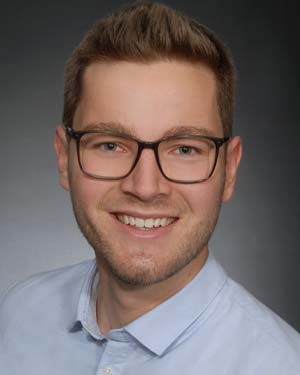}}]{Oliver Schweins} received the bachelor’s (Hons.) and master's degree (Hons.) in electrical engineering  from Paderborn University, Paderborn, Germany, in 2024 and 2026, respectively.

He has been working as a Student Assistant with the Department of Power Electronics and Electrical Drives, Paderborn University, from 2021 until 2026. His research interests include system identification and intelligent control methods.
\end{IEEEbiography}

\begin{IEEEbiography}[{\includegraphics[width=1in,height=1.25in,clip,keepaspectratio]{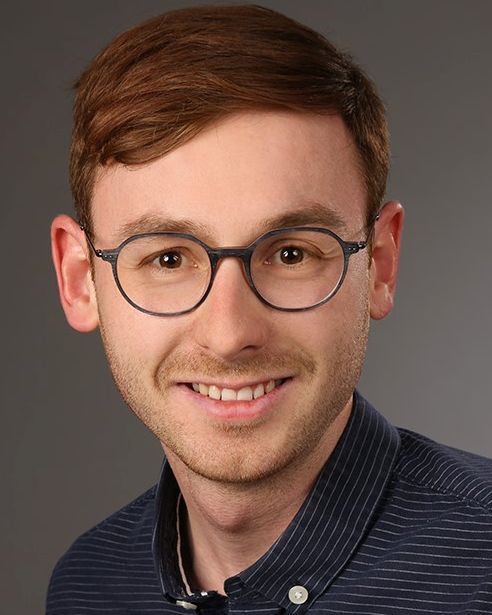}}]{Lukas Hölsch} received his B.Eng. and M.Eng. degrees in electrical engineering from HTWG Konstanz University of Applied Sciences, Germany, in 2020 and 2022 respectively. 

From 2022 to 2024 he worked as a Research Associate for the Department of Power Electronics and Electrical Drives, Paderborn University, Germany. Since then, he is working as a Research Associate at the Chair of Interconnected Automation Systems (IAS), University of Siegen, Siegen, Germany. His research interests include power electronics as well as control of electrical drives, in particular the calculation of optimized pulse patterns for highly utilized permanent magnet synchronous machines.
\end{IEEEbiography}

\begin{IEEEbiography}[{\includegraphics[width=1in,height=1.25in,clip,keepaspectratio]{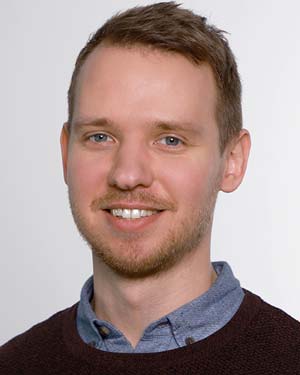}}]{Wilhelm Kirchgässner} received
the B.Sc. degree in electrical engineering from the
OWL University of Applied Sciences and Arts,
Lemgo, Germany, in 2013, and the M.Sc. (Hons.)
and doctoral (Hons.) degrees in electrical engineering
from Paderborn University, Paderborn, Germany,
in 2016 and 2024, respectively.

He is currently with Beckhoff Automation, Verl,
Germany, as a Data Scientist working on planar drive
control and system identification. His research interests
include the application of machine learning
to electrical drive technologies, in particular data-driven thermal modeling of
highly utilized traction drives. 
\end{IEEEbiography}

\begin{IEEEbiography}[{\includegraphics[width=1in,height=1.25in,clip,keepaspectratio]{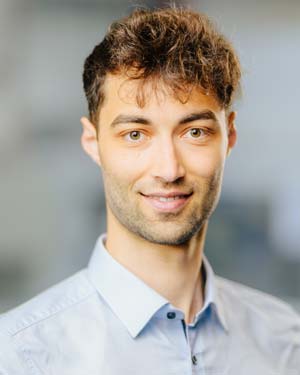}}]{Till Piepenbrock} received the bachelor’s and master’s
(Hons.) degrees in electrical engineering from
Paderborn University, Paderborn, Germany, in 2019
and 2022, respectively.

He is currently working as a Research Associate
with the Department of Power Electronics and Electrical
Drives, Paderborn University. His research interests
include high-frequency dc–dc converter design
with a focus on the optimization of magnetic
components.
\end{IEEEbiography}

\begin{IEEEbiography}[{\includegraphics[width=1in,height=1.25in,clip,keepaspectratio]{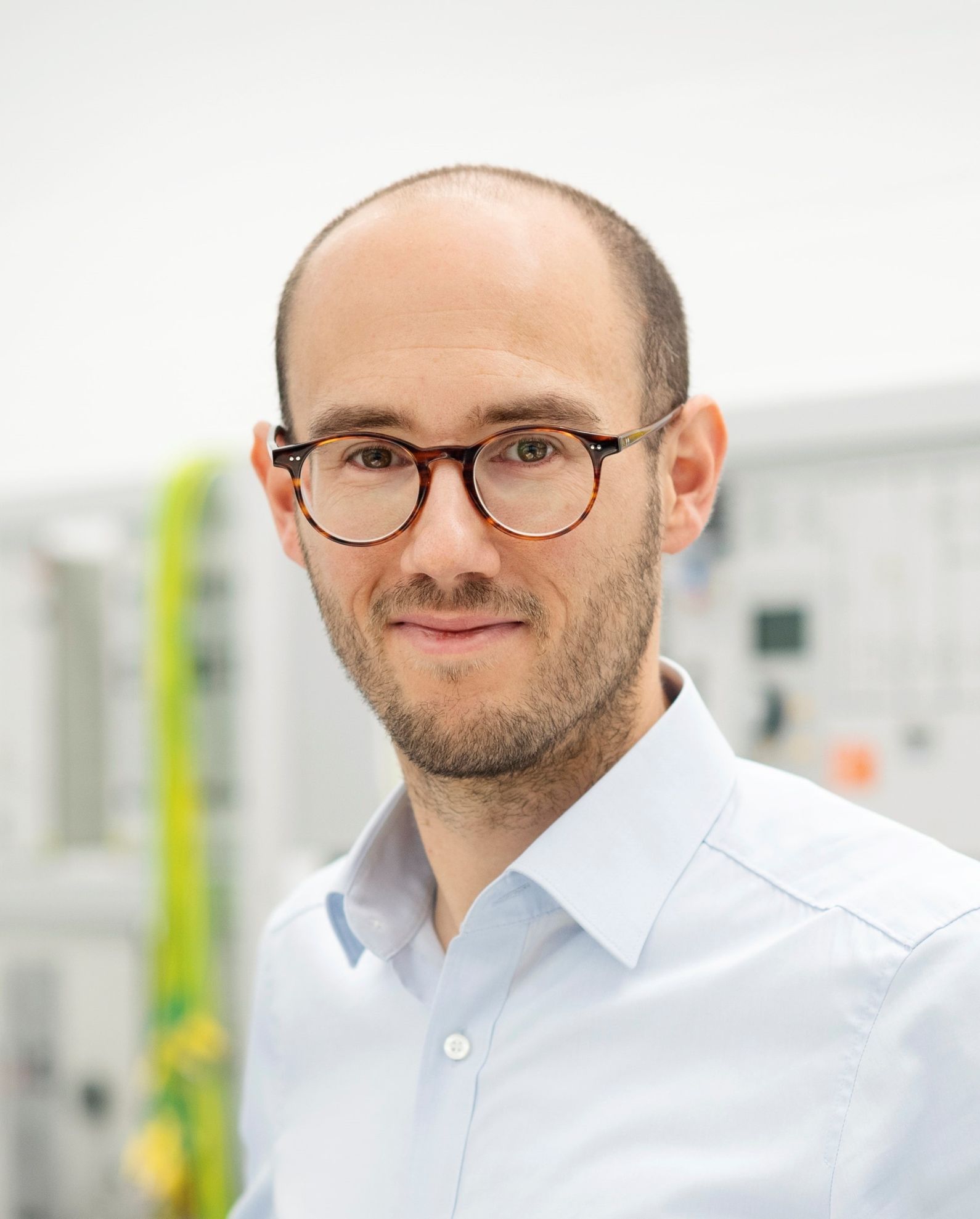}}]{Oliver Wallscheid} (Senior Member, IEEE) received the bachelor’s and master’s degrees (Hons.) in industrial engineering and the doctorate degree (Hons.) in electrical engineering from Paderborn University, Paderborn, Germany, in 2010, 2012, and 2017, respectively. 
    
He has been a senior research fellow with the Department of Power Electronics and Electrical Drives and acting professor of the Automatic Control Department, Paderborn University. Since 2024, he is a full professor at University of Siegen, Siegen, Germany, and heading the Interconnected Automation Systems (IAS) Chair. His research interests include modeling, design, and control of electrical power systems focusing on decentralized grids, power electronics, and electrical drives.
\end{IEEEbiography}

\vfill
\end{document}